\documentclass[12pt]{article}
\usepackage{graphicx}

\newcommand{\lesssim}{\raise.3ex\hbox{$<$\kern-.75em\lower1ex\hbox{$\sim$}}}
\newcommand{\grtssim}{\raise.3ex\hbox{$>$\kern-.75em\lower1ex\hbox{$\sim$}}}

\title{Two lectures on color superconductivity\thanks{Lectures 
delivered at the IARD 2004 conference, Saas Fee, Switzerland,
June 12 - 19, 2004, and at the Helmholtz International Summer 
School and Workshop on {\sl Hot points in Astrophysics and 
Cosmology}, JINR, Dubna, Russia, August 2 - 13, 2004.}}

\author{Igor A. Shovkovy\thanks{On leave from
Bogolyubov Institute for Theoretical Physics, 
03143, Kiev, Ukraine.}\\[3mm]
Frankfurt Institute for Advanced Studies \\
and\\
Institut f\"{u}r Theoretische Physik,\\
Johann Wolfgang Goethe-Universit\"{a}t, \\
D-60054 Frankurt am Main, Germany}

\begin{document}
\maketitle

\begin{abstract}
The first lecture provides an introduction to the physics 
of color superconductivity in cold dense quark matter. The 
main color superconducting phases are briefly described and 
their properties are listed. The second lecture covers recent 
developments in studies of color superconducting phases in 
neutral and $\beta$-equilibrated matter. The properties of 
gapless color superconducting phases are discussed.
\end{abstract}

\newpage

\tableofcontents

\newpage 

\section{Introduction into color superconductivity}

\subsection{Dense baryonic matter}

Almost all matter around us is made of several dozen chemical 
elements and their isotopes. Each element (atom) has a compact 
nucleus, made of protons and neutrons, and a cloud of electrons 
surrounding the nucleus. The proton and the neutron are baryons,
i.e., ``heavy'' particles. They are about two thousand times 
heavier than the electron. (The actual values of the masses 
are $m_{\rm p} \approx 938.3\,\mbox{MeV}/c^2$, $m_{\rm n} \approx 
939.6\,\mbox{MeV}/c^2$, and $m_{\rm e} \approx 0.511\,\mbox{MeV}/c^2$,
see Ref.~\cite{HDG}.) Thus, the mass of an atom comes mostly from 
its heavy nucleus. A typical size of a nucleus is about five orders 
of magnitude smaller than a typical atomic size. It is clear, 
therefore, that the density of matter inside a nucleus is much 
larger than the density of matter made of chemical elements. In 
fact, typical densities inside nuclei are of order $10^{14}\,\mbox{g}/
\mbox{cm}^{3}$. To get an impression how large this is, it suffices 
to say that a tablespoon of a material with this density would weigh 
about a billion tonnes. Creating even a tiny macroscopic sample of 
such matter in laboratory is impossible. On the other hand, baryonic 
matter at even higher densities exists in the Universe in the 
interior of compact stars. Therefore, it is of fundamental 
importance to study such very dense matter.

Protons and neutrons, as well as all mesons and baryons 
that may appear in dense matter, are strongly interacting particles. 
Their properties and the properties of dense baryonic matter 
are, in principle, described by the microscopic theory of strong 
interactions, Quantum Chromodynamics (QCD). The only problem is that 
QCD is a rather complicated non-Abelian gauge theory. The Lagrangian 
density of the theory is written in terms of quark fields which carry
color charges. Because of the property of confinement, however, 
quarks cannot exist in vacuum as free particles. Instead, they 
combine with other quarks and antiquarks to form color neutral 
hadrons (i.e., mesons and baryons) whose description is not easy
in the framework of QCD. 

It is known that baryons are not point-like particles. They have 
a typical size of about $1\,\mbox{fm}=10^{-13}\,\mbox{cm}$. In 
sufficiently dense matter, therefore, baryons can be forced to stay 
so close to one another that they would overlap. At such densities, 
constituent quarks are shared by neighboring baryons and, with 
increasing the density further, quarks should eventually become 
mobile over large distances. This means that quarks become deconfined. 
When this happens, it does not make sense to talk about hadronic 
matter any more. It is transformed into quark matter. The properties 
of such matter is the topic of these lectures.

It was suggested long time ago that quark matter may exist inside 
central regions of compact stars \cite{quark-star}. By making use of 
the property of asymptotic freedom in QCD \cite{GWP}, it was argued 
that quarks interact weakly, and that realistic calculations taking 
full account of strong interactions are possible for sufficiently 
dense matter \cite{ColPer}. The argument of Ref.~\cite{ColPer} 
consisted of the two main points: (i) the long-range QCD interactions 
are screened in dense medium causing no infrared problems, and
(ii) at short distances, the interaction is weak enough to allow 
the use of the perturbation theory. As we shall see below, the 
real situation in dense quark matter is slightly more subtle.

\subsubsection{Weakly interacting quark matter}

By assuming that very dense matter is indeed made of weakly
interacting quarks, one could try to understand the thermodynamic
properties of the corresponding ground state by first completely 
neglecting the interaction between quarks. In order to construct 
the ground state,
it is important to keep in mind that quarks are fermions, i.e., 
particles with a half-integer spin, $s=1/2$. Therefore, quarks 
should obey the Pauli exclusion principle which says that no two 
identical quarks can simultaneously occupy the same quantum state. 

In the ground state of zero temperature non-interacting
quark matter, quarks occupy all available quantum states with the 
lowest possible energies. This is described formally by the following 
quark distribution function:
\begin{equation}
f_F(\mathbf{k}) =\theta\left(\mu-E_\mathbf{k}\right), \quad \mbox{at} 
\quad T=0,
\label{f_F_T=0}
\end{equation}
where $\mu$ is the quark chemical potential, and $E_\mathbf{k}\equiv
\sqrt{k^2+m^2}$ is the energy of a free quark (with mass $m$) in the 
quantum state with the momentum $\mathbf{k}$ (by definition, $k\equiv 
|\mathbf{k}|$). As one can see, $f_F(\mathbf{k})=1$ for the states 
with $k<k_F\equiv \sqrt{\mu^2-m^2}$, indicating that all states with 
the momenta less than the Fermi momentum $k_F$ are occupied. The 
states with the momenta greater than the Fermi momentum $k_F$ are 
empty, i.e., $f_F(\mathbf{k})=0$ for $k>k_F$. 

In the chiral limit, i.e., when the quark mass is vanishing, the 
pressure of zero temperature free quark matter is given by 
\cite{Kapusta}
\begin{equation}
P^{(0)} = 2 N_f N_c \int \frac{d^3\mathbf{k}}{(2\pi)^3}
\left(\mu-E_\mathbf{k}\right) \theta\left(\mu-E_\mathbf{k}\right) 
-B =\frac{N_f N_c}{12 \pi^2} \mu^4 - B,
\label{pressure_0}
\end{equation}
where the overall factor $2 N_f N_c$ counts the total number of
degenerate quark states, namely 2 spin states (i.e., $s=\pm 1/2$), 
$N_f$ flavor states (e.g., up, down and strange), and $N_c$ color 
states (e.g., red, green and blue). The extra term $B$ in the 
expression for the pressure (\ref{pressure_0}), called the bag 
constant, was added by hand. This term effectively assigns a 
nonzero contribution to the vacuum pressure and, in this way, 
provides the simplest modelling of the quark confinement in QCD 
\cite{bag-model}. 
[Note that the vacuum pressure is higher than the pressure of 
quark matter when $\mu<\left(4\pi^2 B/N_f\right)^{1/4}$.] Here,
we ignore the requirements of the charge neutrality and the 
$\beta$ equilibrium in quark matter. These will be addressed 
in detail in the second lecture. The energy density of quark 
matter reads
\begin{equation}
\epsilon^{(0)} \equiv \mu \frac{\partial P^{(0)}}{\partial \mu} 
- P^{(0)} =\frac{N_f N_c}{4 \pi^2} \mu^4 +B.
\label{e_0}
\end{equation}
Thus, one arrives at the following equation of state:
\begin{equation}
P^{(0)}=\frac{1}{3} \left(\epsilon^{(0)} -4B\right).
\label{EoS_0}
\end{equation}
This equation of state could be easily generalized to the case of
nonzero quark masses and/or to the case of non-equal chemical 
potentials for different quarks. Also, it could be further improved 
by adding the lowest order corrections due to the interaction. 
For example, to leading order in coupling, the interaction results 
in the following correction to the pressure \cite{FreMcL}:
\begin{equation}
\delta P^{(0)} =-\frac{\alpha_s N_f (N_c^2-1)}{16 \pi^3} \mu^4 
+O\left(\alpha_s^2\right),
\label{pressure_cor}
\end{equation}
where $\alpha_s\equiv g^2/4\pi$ is the value of the running coupling 
constant of strong interactions defined at the scale of the quark 
chemical potential. The next to leading order corrections can also
be calculated \cite{FreMcL}.

It is generally believed that the equation of state in 
Eq.~(\ref{pressure_0}), or its generalization, provides a good 
approximation for the description of weakly interacting quark 
matter. Below we argue, however, that some underlying assumptions 
regarding the ground state of such quark matter are not correct. 

\subsection{Cooper instability and color superconductivity}

It appears that the perturbative ground state of quark matter,
characterized by the distribution function in Eq.~(\ref{f_F_T=0}),
is unstable when there is an attractive (even arbitrarily weak in 
magnitude\,!) interaction between quarks. This is because of the 
famous Cooper instability \cite{Cooper} that develops as a result
of the formation of Cooper pairs $\langle q_{\mathbf{k}}\, 
q_{-\mathbf{k}}\rangle$ made of quarks from around the highly 
degenerate Fermi surface, i.e., quarks with the absolute value of 
momenta $k\simeq k_F$. 
Cooper pairs are bosons, and they occupy the same lowest energy 
quantum state at zero temperature, producing a Bose condensate. 
In the presence of such a condensate of Cooper pairs, 
the ground state of quark matter becomes a (color) superconductor. 
This is very similar to the ground state of the electron system 
in the Bardeen-Cooper-Schrieffer theory of low-temperature 
superconductivity \cite{BCS}. The only real difference comes from 
the fact that quarks, unlike electrons, come in various flavors 
(e.g., up, down and strange) and carry non-Abelian color charges. 
To emphasize this difference, superconductivity in quark matter 
is called {\it color superconductivity}. For recent reviews on 
color superconductivity, see Ref.~\cite{reviews}.

It was known for a long time that dense quark matter should be a 
color superconductor \cite{BarFra,Bail}. In many studies, however,
this fact was commonly ignored. Only recently, the potential importance 
of this phenomenon was appreciated. To large extent, this was triggered 
by the observation in Ref.~\cite{cs} that the value of the color 
superconducting gap could be as large as $100\,\mbox{MeV}$ at baryon 
densities existing in the central regions of compact stars, i.e., at
densities which are a few times larger than the normal nuclear 
density, $n_0\simeq 0.15 \mbox{~fm}^{-3}$. This very natural estimate 
for the value of the gap in QCD, in which a typical energy scale itself 
is $200\,\mbox{MeV}$, opened a wide range of new theoretical possibilities, 
and the subject bursted with numerous studies. The main reason is that the 
presence of such a large energy gap in the quark spectrum may allow 
to extract clear signatures of color superconducting states 
of matter in observational data from compact stars. 

As in low-temperature superconductors, one of the main consequences 
of color superconductivity in dense quark matter is the appearance 
of a nonzero energy gap in the one-particle spectrum,
\begin{equation}
{\cal E}_{\mathbf{k}} 
= \sqrt{\left(E_{\mathbf{k}}-\mu\right)^2+\Delta^2},
\label{E_qp}
\end{equation}
where $\Delta$ is the gap. The presence of the gap in the energy 
spectrum should affect transport properties (e.g., conductivities 
and viscosities) of quark matter. Thus, if quark matter exists in the 
interior of compact stars, this will be reflected, for example, in the 
cooling rates and in the rotational slowing down of such stars. Also, a
nonzero gap modifies thermodynamic properties, e.g., the specific 
heat and the equation of state. In application to stars, this could
modify theoretical predictions for the mass-radius relations, or even 
suggest the existence of a new family of compact stars.

Color superconductivity may affect directly as well as indirectly 
many other observed properties of stars. In some cases, for example, 
superconductivity may be accompanied by the baryon superfluidity 
and/or the electromagnetic Meissner effect. If matter is superfluid,
rotational vortices would be formed in the stellar core, and they
would carry a portion of the angular momentum of the star. Because 
of the Meissner effect, the star interior could become threaded 
with magnetic flux tubes. In either case, the star evolution may 
be affected. 

In general, it is of great phenomenological interest to perform a 
systematic study of all possible effects of color superconductivity 
in compact stars. Before this can be done, however, one needs to 
know the structure of the QCD phase diagram and properties of 
various color superconducting phases in detail. Despite the 
recent progress in the field, such knowledge still remains 
patchy. While many different phases of quark matter have been 
proposed, there is no certainty that all possibilities have 
already been exhausted. This is especially so when additional 
requirements of charge neutrality and $\beta$ equilibrium are 
imposed. 

In the rest of this lecture, the main color superconducting phases,
that involve different number of pairing quark flavors, will be 
reviewed briefly. The role of charge neutrality and $\beta$ 
equilibrium in quark matter will be discussed in the second 
lecture.

\subsection{Two-flavor color superconductivity ($N_f=2$)}

It is instructive to start this lecture from the description of the 
simplest color superconducting phase, namely the two-flavor color 
superconductor (2SC). This is a color superconducting phase with 
the spin-0 Cooper pairing in quark matter made of up and down 
quarks. 

As we discussed in the previous subsection, color superconductivity 
comes as a result of the Cooper instability driven by an attractive 
interaction between quarks. To study this instability in detail, 
one needs to specify the origin and the strength of the interaction.
At very large densities, when quark matter is weakly interacting,
the dominant interaction between quarks is provided by the one-gluon
exchange. Then, the Cooper pairing dynamics, including the screening 
effects due to dense medium as well as the effects due to higher 
order corrections, can be studied from first principles within the 
framework of QCD. Some key details of the analysis in this regime 
are given below.

We consider quark fields $\psi_i^a$ in the fundamental representation 
of the SU(3)$_c$ color gauge group. These fields carry flavor ($i=1,2$) 
and color ($a=1,2,3$) indices. The QCD Lagrangian density reads
\begin{equation}
{\cal L}_{\rm QCD} = \bar\psi_i^a\left(i\gamma^\mu \partial_\mu
+\gamma^0\mu -m^{(0)}_i\right) \psi_i^a 
+g A^{A}_{\mu} \bar\psi_i^a \gamma^\mu T^{A}_{ab} \psi_i^b
-\frac{1}{4}G^{A}_{\mu\nu}G^{A,\mu\nu},
\label{QCD-action}
\end{equation}
where $A^{A}_{\mu}$ is the vector gauge field in the adjoint 
representation of SU(3)$_c$, $G^{A}_{\mu\nu}= \partial_\mu A^{A}_{\nu}
-\partial_\nu A^{A}_{\mu}+g f^{ABC}A^{B}_{\mu}A^{C}_{\nu}$ is the field
strength, and the generators of color transformations are defined as
$T^{A}_{ab} \equiv \frac{1}{2}\left(\lambda^{A} \right)_{ab}$ where 
$\lambda^{A}$ are the Gell-Mann matrices. The current quark masses
and the quark chemical potential are denoted by $m^{(0)}_i$ and $\mu$,
respectively. At high densities, one can neglect small current masses 
of quarks. Then, the QCD Lagrangian density (\ref{QCD-action}) becomes 
invariant under SU(2)$_L\times$SU(2)$_R$ global chiral transformations. 

In QCD, the constituent masses of quarks $m_i$ are generated dynamically,
and they can be very different from the current masses $m^{(0)}_i$, 
appearing in the Lagrangian density. At zero density, for example, 
typical values of the constituent quark masses are of order $\frac{1}{3}
m_{\rm n}\approx 313\,\mbox{MeV}$ even in the chiral limit. At high 
densities, on the other hand, the masses of the up and down quarks
become small. This is because they are proportional to the value of the 
chiral condensate $\langle \bar\psi_L \psi_R\rangle$ which melts in dense 
matter. Thus, it is often also justified to neglect the constituent quark 
masses in studies of color superconducting phases. 

\begin{figure}[ht]
\begin{center}
\includegraphics[width=0.4\textwidth]{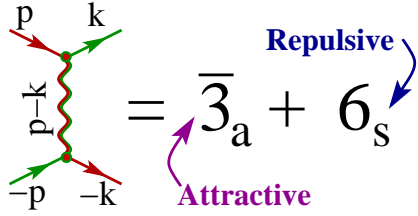}
\caption{The diagrammatic representation of the one-gluon 
exchange interaction between two quarks in QCD. The color 
structure of the corresponding amplitude contains an 
antisymmetric antitriplet and a symmetric sextet channel.}
\label{fig-1-glu}
\end{center}
\end{figure}

The interaction vertex in the Lagrangian density (\ref{QCD-action}) has 
a non-trivial color structure given by the color generators $T^{A}_{ab}$. 
As a result, the quark-quark scattering amplitude in the one-gluon 
exchange approximation is proportional to the following color tensor:
\begin{equation}
\sum_{A=1}^{N_c^2-1} T^{A}_{a a^\prime} T^{A}_{b^\prime b} =
-\frac{N_c+1}{4N_c}\left(\delta_{a a^\prime}\delta_{b^\prime b}
-\delta_{ab^\prime}\delta_{a^\prime b}\right)
+\frac{N_c-1}{4 N_c}\left(\delta_{a a^\prime}\delta_{b^\prime b}
+\delta_{a b^\prime}\delta_{a^\prime b}\right).
\label{T-T}
\end{equation}
The first antisymmetric term corresponds to the {\em attractive} 
antitriplet channel, while the second symmetric term corresponds 
to the repulsive sextet channel, see Fig.~\ref{fig-1-glu}.
It is the first antisymmetric antitriplet channel that plays the 
crucial role in Cooper pairing.

Here, it is appropriate to note that quark matter is unlikely to be 
truly weakly interacting at moderate densities existing in the central 
regions of compact stars, i.e., $n\lesssim 10n_0$. In this case, 
the use of the microscopic theory of strong interactions is very 
limited, and one has to rely on various effective models of QCD. A 
very simple type of such a model is the Nambu-Jona-Lasinio (NJL) model 
with a local four-fermion interaction \cite{NJL}. Originally, this 
model was introduced for the description of nucleons with dynamically 
generated masses. Nowadays, it is also commonly used for the description 
of quarks. For example, the NJL type models were used in 
Refs.~\cite{cs,4fermi,4fermi-rev,huang_2sc} for the description of color 
superconducting phases. One of the simplest NJL models, that respects 
the SU(2)$_L\times$SU(2)$_R$ global chiral symmetry (in the limit 
$m^{(0)}_i\to 0$), is defined by the following Lagrangian density 
\cite{SKP}:
\begin{eqnarray}
{\cal L}_{\rm NJL} &=& \bar\psi_i^a\left(i\gamma^\mu \partial_\mu
+\gamma^0\mu -m^{(0)}_i\right) \psi_i^a 
+ G_S\left[(\bar\psi \psi)^2 
+ (i\bar\psi \gamma_5\vec{\tau}\psi)^2\right]
\nonumber\\
&+&G_D (i \bar{\psi}^C \varepsilon \epsilon^a \gamma_5 \psi )
 (i \bar{\psi} \varepsilon \epsilon^a \gamma_5 \psi^C),
\label{NJL-action}
\end{eqnarray}
where $\psi^C=C \bar{\psi}^T$ is the charge-conjugate spinor and 
$C=i\gamma^2\gamma^0$ is the charge conjugation matrix. The matrix 
$C$ is defined so that $C\gamma_{\mu} C^{-1}=-\gamma_{\mu}^{T}$. 
Regarding the other notation, $\vec{\tau}=(\tau^1,\tau^2,\tau^3)$ are 
the Pauli matrices in the flavor space, while $(\varepsilon)^{ik} \equiv 
\varepsilon^{ik}$ and $(\epsilon^a)^{bc} \equiv \epsilon^{abc}$ are the 
antisymmetric tensors in the flavor and in the color spaces, respectively. 
The dimensionful coupling constant $G_S=5.01\,\mbox{GeV}^{-2}$ and the 
momentum integration cutoff parameter $\Lambda=0.65\,\mbox{GeV}$ (which 
appears only in loop calculations) are adjusted so that the values 
of the pion decay constant and the value of the chiral condensate
take their standard values in vacuum QCD: $F_\pi = 93$ MeV and 
$\langle\bar{u}u\rangle=\langle\bar{d}d\rangle =(-250\,\mbox{MeV})^3$
\cite{SKP}. The strength of the coupling constant $G_D$ is taken to be 
proportional to the value of $G_S$ as follows: $G_D=\eta G_S$ where 
$\eta$ is a dimensionless parameter of order 1. Note that the value of 
$\eta$ is positive which corresponds to having the antisymmetric diquark 
channel {\em attractive}. This is motivated by the microscopic QCD 
interaction, as well as by the instanton motivated models \cite{cs}.

 \subsubsection{Color and flavor structure of the condensate}

Now let us discuss the color and flavor structure of the condensate 
of Cooper pairs in the 2SC ground state. This can be determined from
the nature of the diquark interaction and the Pauli exclusion principle. 

The color structure of the one-gluon exchange interaction in 
Eq.~(\ref{T-T}) shows that the color antisymmetric antitriplet 
channel is attractive. The antisymmetric color channel is also 
attractive in the instanton motivated models of Ref.~\cite{cs}.
This is reflected in the color structure of the condensate of 
spin-0 Cooper pairs,
\begin{equation}
\left\langle \left(\bar{\psi}^C\right)_i^a \gamma^5 \psi_j^b 
\right\rangle
\sim \varepsilon_{ij} \epsilon^{abc} ,
\label{2sc-cond}
\end{equation}
which is antisymmetric in the color indices of the constituent
quarks. As one can check (not shown explicitly), this spin-0 
condensate is also antisymmetric in Dirac indices. Then, in 
accordance with the Pauli exclusion principle, it should be 
antisymmetric in flavor indices as well. The presence of the 
$\gamma^5$ matrix on the left hand side of Eq.~(\ref{2sc-cond})
makes the condensate a scalar. A similar condensate without 
$\gamma^5$ is a pseudoscalar which is disfavored in QCD because 
of the instanton effects. 

The color antitriplet condensate (\ref{2sc-cond}) may have an 
arbitrary orientation in the color space. By making use of the 
global color transformations, this orientation can be changed 
as convenient. It is conventional to choose the condensate to 
point in the third (blue) color direction, $\left\langle 
\left(\bar{\psi}^C\right)_i^a \gamma^5 \psi_j^b \right\rangle
\sim \varepsilon_{ij} \epsilon^{ab3}$. In this case, the Cooper 
pairs in the 2SC phase are made of the red and green quarks only, 
and the blue quarks do not participate in the pairing. These 
unpaired blue quarks give rise to gapless quasiparticles in the 
low-energy effective theory. 

The flavor antisymmetric structure in Eq.~(\ref{2sc-cond}) corresponds 
to a singlet representation of the global SU(2)$_L\times$SU(2)$_R$ 
chiral group. This means that the chiral symmetry is not broken in 
the 2SC ground state. In fact, there are no other global continuous 
symmetries which are broken in the 2SC phase. There exist, however, 
several approximate symmetries which are broken. One of them is the 
approximate U(1)$_A$ symmetry which is a good symmetry at high density 
when the instantons are screened \cite{scr_inst}. Its breaking in the
2SC phase results in a pseudo-Nambu-Goldstone boson \cite{low-e-2sc}. 
Additional four exotic pseudo-Nambu-Goldstone states may appear as a 
result of a less obvious approximate axial color symmetry discussed 
in Ref.~\cite{pseudoNG}.

In the 2SC ground state, the vector-like SU(3)$_c$ color gauge group 
of QCD is broken down to the SU(2)$_c$ subgroup. Therefore, five out 
of total eight gluons of SU(3)$_c$ become massive due to the 
Anderson-Higgs mechanism. The other three gluons, which correspond 
to the unbroken SU(2)$_c$, do not interact with the gapless blue 
quasiparticles. They give rise to a pure SU(2)$_c$ gluodynamics. 
The red and green quasiparticles decouple from this low-energy 
SU(2)$_c$ gluodynamics because they are gapped \cite{RSS}. 

 \subsubsection{Quark propagator in the Nambu-Gorkov basis}

To simplify the study of diquark condensates in dense quark matter,
instead of using the Dirac spinors for quark fields, it is more
convenient to introduce the following eight-component Nambu-Gorkov 
spinors:
\begin{equation}
\Psi \equiv \left(\begin{array}{l}\psi \\ \psi^C \end{array}\right),
\end{equation}
where $\psi^C$ is the charge-conjugate spinor, defined earlier.
The structure of the inverse quark propagator in this basis reads
\begin{equation}
S_0^{-1} = \left(
\begin{array}{cc}
\left[ G_0^+ \right]^{-1} & 0 \\
0 & \left[ G_0^- \right]^{-1}
\end{array}
\right),
\end{equation}
where
\begin{equation}
\left[ G_0^\pm \right]^{-1} = \gamma^\mu K_\mu \pm \mu \gamma_0
\end{equation}
are the inverse Dirac propagators for massless quarks ($G_0^{+}$) 
and charge-conjugate quarks ($G_0^{-}$). 

In the ground state with Cooper pairing, see Eq.~(\ref{2sc-cond}),
the quark propagator in the Nambu-Gorkov basis also has nonzero 
off-diagonal components,
\begin{equation}
S^{-1} = \left(
\begin{array}{cc}
\left[ G_0^+ \right]^{-1} & \Delta^{-} \\
\Delta^{+} & \left[ G_0^- \right]^{-1}
\end{array}
\right),
\label{inv_prop}
\end{equation}
where $\Delta^{-} = -i \epsilon^{3}\varepsilon\gamma^5 \Delta$ and
$\Delta^{+} \equiv \gamma^0 \left(\Delta^{-}\right)^{\dagger} \gamma^0
= -i \epsilon^{3}\varepsilon\gamma^5\Delta^{*}$ are the matrices in 
the Dirac space, and $\Delta$ is the diquark gap parameter. This is 
the simplest ansatz for the quark propagator in the color superconducting
ground state. It can be made more general, for example, by adding a
regular (diagonal) part of the quark self-energy. Such a self-energy 
plays an essential role in the dynamics of chiral symmetry breaking. 
In dense quark matter, on the other hand, it is not so important. 
Thus, for simplicity, it is neglected here (the effects of the diagonal 
part of the self-energy were studied in Ref.~\cite{self-e}). 

From the inverse propagator in Eq.~(\ref{inv_prop}), one can derive 
the following expression for the quark propagator:
\begin{equation}
S = \left(\begin{array}{cc}
G^{+} & \Xi^{-} \\
\Xi^{+} & G^{-}
\end{array}\right),
\label{quarkpropagator}
\end{equation}
where
\begin{eqnarray}
G^{\pm} &=&
\left[\left(G_0^{\pm}\right)^{-1}
-\Delta^{\mp}G_0^{\mp}\Delta^{\pm}\right]^{-1}, \\
\Xi^{\pm} &=& -G_0^{\mp} \Delta^{\pm} G^{\pm} .
\end{eqnarray}
This structure of the quark propagator carries basic
information about the properties of the 2SC ground state.
For instance, the location of its poles in the energy plane
gives the quasiparticle dispersion relations.

The only quantity that remains unknown in the ansatz (\ref{inv_prop}) 
is the value of the gap parameter $\Delta$. This should be determined 
from a self-consistent gap equation that takes into account the pairing 
dynamics between quarks. This is discussed in more detail in the next 
three subsections. 

 \subsubsection{One-gluon exchange interaction}

In dense QCD, in which quarks are weakly interacting, the one-gluon
exchange is the dominant interaction between quarks. It appears, 
however, that the one-gluon interaction is partially screened by 
surrounding dense medium. Thus, before studying the pairing 
dynamics between quarks, we first address the modification of 
gluon properties due to dense medium. 

The inverse propagator of the medium modified gluon can be written 
in the following form: 
\begin{equation}
\left({\cal D}^{-1}\right)_{\mu\nu}^{AB} = 
\left(D_0^{-1}\right)_{\mu\nu}^{AB}
+\Pi_{\mu\nu}^{AB},
\label{glu_prop}
\end{equation}
where $\Pi_{\mu\nu}^{AB}$ is the gluon self-energy, or the polarization 
tensor. The gluons with soft momenta, $p\lesssim g_s\mu$, play the key 
role in Cooper pairing of quarks. The main contribution to the 
corresponding soft gluon polarization tensor comes from the quark loop 
with hard internal momenta of order $\mu$. This contribution is large
compared to all other contributions in QCD (e.g., those from the
ghost and the gluon loops) because it is proportional to the density 
of quark states at the Fermi surface (as well as to the coupling 
constant $\alpha_s$), i.e., $\Pi_{\mu\nu}^{AB}\sim 
\alpha_s\mu^2$. The corresponding approximation, in which the
sub-leading terms suppressed by powers of the quark chemical
potential $\mu$ are neglected, is called the hard dense loop 
(HDL) approximation. 

The calculation of the polarization tensor in the HDL approximation 
was performed in Refs.~\cite{Vija,Manuel} (in the case of dense QCD 
in $2+1$ dimensions, similar calculations were performed in 
Ref.~\cite{qcd2+1}). Here, we present the final result and briefly 
comment on its main features. 

The HDL polarization tensor can be written as follows: 
$\Pi_{\mu\nu}^{AB}\equiv \delta^{AB}\Pi_{\mu\nu}$, where 
the explicit form of the components of $\Pi_{\mu\nu}$ are 
given by
\begin{eqnarray}
\Pi^{00}(p_0, \mathbf{p}) & = & \Pi_{l}(p_0, \mathbf{p}) ,
\label{Pi-00}\\
\Pi^{0i}(p_0, \mathbf{p}) & = &
p_0 \frac{p^i}{p^2} \Pi_{l} (p_0, \mathbf{p}) , 
\label{Pi-0i}\\
\Pi^{ij}(p_0, \mathbf{p}) & = & \left
( \delta^{ij}- \frac{p^i p^j}{p^2} \right)
\Pi_{t} (p_0,\mathbf{p})+ \frac{p^i p^j} {p^2}
\frac{p_0^2}{p^2} \Pi_{l} (p_0, \mathbf{p}),
\label{Pi-ij}
\end{eqnarray}
with
\begin{eqnarray}
\Pi_{l}(p_0,\mathbf{p})&=&m_D^2\left(\frac{p_0}{2p}
\ln\left|\frac{p_0+p}{p_0-p}\right|-1
-i\pi\frac{p_0}{2p}\theta(-p^2) \right),\\
\Pi_{t}(p_0,\mathbf{p})&=&\frac{1}{2}m_D^2-\frac{P^2}{2p^2}
\Pi_{l}(p_0,\mathbf{p}).
\end{eqnarray}
Here, we use the notation $p^2\equiv \mathbf{p}^2$, $P^2=p_0^2-p^2$, 
and $m_D^2\equiv 2N_f\alpha_{s}\mu^2/\pi$. It is remarkable that
this result coincides with the polarization tensor derived in the 
framework of the {\em classical} transport theory of dense Yang-Mills 
plasma \cite{Heinz}. One can check that the HDL polarization tensor
is transverse, i.e., 
\begin{equation}
P^{\mu} \Pi_{\mu\nu}(p_0, \mathbf{p})=0.
\end{equation}
It is convenient to write the tensor $\Pi_{\mu\nu}$ in terms of two
transverse projection operators,
\begin{equation}
\Pi_{\mu\nu}=-O^{(1)}_{\mu\nu} \Pi_{t}
+O^{(2)}_{\mu\nu} \left(2\Pi_{t}-m_D^2\right).
\end{equation}
where
\begin{eqnarray}
O^{(1)}_{\mu\nu}&=& g_{\mu\nu}-u_{\mu} u_{\nu}
+\frac{\mathbf{p}_{\mu}\mathbf{p}_{\nu}}{p^{2}},
\label{def-O1} \\
O^{(2)}_{\mu\nu}&=& u_{\mu} u_{\nu}
-\frac{\mathbf{p}_{\mu}\mathbf{p}_{\nu}}{p^{2}}
-\frac{P_{\mu}P_{\nu}}{P^{2}},
\label{def-O2} 
\end{eqnarray}
and $u_{\mu}=(1,0,0,0)$ is the time-like four-vector that 
defines the rest frame of dense quark medium. In order to have 
a complete set of projectors such that $\sum_I O^{(I)}_{\mu\nu}
=g_{\mu\nu}$, one has to include also the following longitudinal 
projector:
\begin{equation}
O^{(3)}_{\mu\nu}= \frac{P_{\mu}P_{\nu}}{P^{2}}. 
\label{def-O3} 
\end{equation}
By making use of this set of three projectors, inverse gluon propagator 
in Eq.~(\ref{glu_prop}) can be represented as follows:
\begin{equation}
\left({\cal D}^{-1}\right)_{\mu\nu}^{AB}
=i\delta^{AB} \left[\left(P^2-\Pi_{t}\right) O^{(1)}_{\mu\nu}
+\left(P^2+2\Pi_{t}-m_D^2\right) O^{(2)}_{\mu\nu}
+\frac{P^2}{\lambda} O^{(3)}_{\mu\nu}\right],
\label{inv-D}
\end{equation}
where $\lambda$ is the gauge fixing parameter. This expression can
be easily inverted, leading to the following result for the gluon
propagator:
\begin{equation}
{\cal D}_{\mu\nu}^{AB} =-i\delta^{AB}\left(
\frac{1}{P^2-\Pi_{t}} O^{(1)}_{\mu\nu}
+\frac{1}{P^2+2\Pi_{t}-m_D^2} O^{(2)}_{\mu\nu}
+\frac{\lambda}{P^2} O^{(3)}_{\mu\nu}\right).
\label{D}
\end{equation}
The gluon modes which are transverse and longitudinal with respect to 
the {\em three-momentum} $\mathbf{p}$ are called the magnetic and electric 
modes, respectively. These two types of physical gluon modes are 
described by the following fields:
\begin{eqnarray}
a^{\rm (mag)}_{\mu}&=& O^{(1)}_{\mu\nu}A^{\nu},
\label{def-mag} \\
a^{\rm (el)}_{\mu}&=&O^{(2)}_{\mu\nu}A^{\nu} .
\label{def-el} 
\end{eqnarray}
From Eq.~(\ref{D}) one can see that the propagators of these modes 
are naturally separated from each other, as well as from the unphysical 
longitudinal mode, described by the field $a^{\rm (\parallel)}_{\mu}
=O^{(3)}_{\mu\nu}A^{\nu}$. 

In the study of the Cooper pairing dynamics, the most relevant 
part of the one-gluon exchange interaction is the interaction 
with the space-like momenta such that $p_0\ll p$. This should 
be clear from the kinematics of the quark scattering around the 
Fermi surface: the energy exchange between pairing quarks
is typically much smaller than the change of their momenta.
It is justified, therefore, to use the approximate form of 
$\Pi_{t}$, obtained in the region $p_0\ll p$,
\begin{equation}
\Pi_{t}\simeq m_D^2\left(i\pi \frac{p_0}{4p}
\theta(P^2)+\frac{p_0^2}{p^2}+\dots\right).
\label{asym-2}
\end{equation}
By substituting this asymptote into the gluon propagator (\ref{D}), 
one can check that the magnetic modes produce a long-range (dynamically 
screened) interaction, while the electric mode works only at 
short distances, $r \lesssim 1/m_D$. 

In the next subsection, the HDL gluon propagator is used in the study 
of the Cooper pairing dynamics. As we shall see there, the long-range 
interaction mediated by the magnetic modes will play a particularly
important role in the dynamics.

 \subsubsection{Gap equation in QCD}

Color superconductivity is an essentially non-perturbative phenomenon
which cannot be addressed with perturbative techniques even if the 
theory is weakly interacting. The standard method for studying
superconductivity is the method of the Schwinger-Dyson equations. 
In general, the complete set of Schwinger-Dyson equations contains 
an infinite number of coupled equations for Green's functions. There
exist no systematic ways of solving such equations exactly. However, 
there exist various standard approximations that can be justified if
the dynamics is controlled by a small parameter (e.g., a weak coupling 
constant, or a small $1/N$ parameter where $N$ is the number of flavors 
or colors in the model). 

\begin{figure}[ht]
\begin{center}
\includegraphics{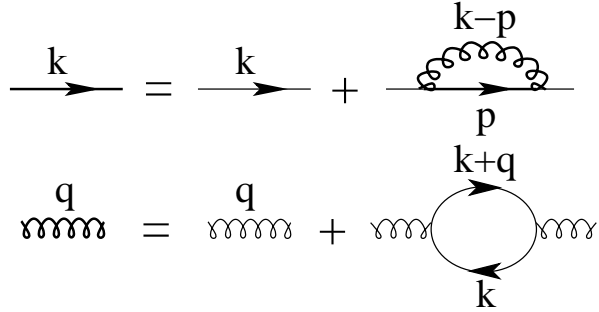}
\caption{The graphical representation of the Schwinger-Dyson
equation in the improved rainbow approximation. The quark propagator
in the Nambu-Gorkov basis and the gluon propagator are denoted by 
the solid lines and the wavy lines, respectively.}
\label{fig-sd-eq}
\end{center}
\end{figure}

\begin{figure}[ht]
\begin{center}
\includegraphics[width=0.5\textwidth]{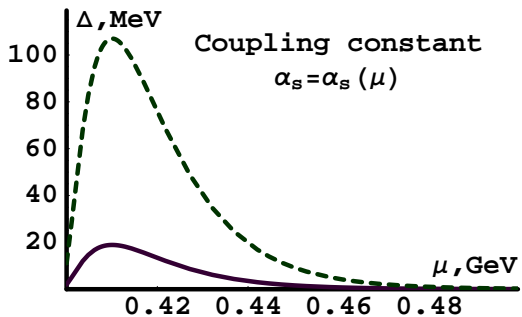}
\caption{The value of the gap from Eq.~(\ref{gap-sol}) as a function 
of the quark chemical potential. The coupling constant is defined at 
the scale of the quark chemical potential. The solid (dashed) line 
gives the result with (without) taking the quark wave function 
renormalization into account.}
\label{fig-gap-sol}
\end{center}
\end{figure}

In the case of QCD at asymptotic densities, the main features of 
its dynamics are well described by the so-called improved rainbow 
approximation. In this approximation, one uses the bare quark-gluon 
vertices in the Schwinger-Dyson equation for the quark propagator. 
One also uses the gluon propagator with the screening effects taken 
into account in the HDL approximation. (Note that, in the simple 
rainbow approximation, the screening is usually neglected.) The 
graphical representation of the Schwinger-Dyson equation is shown 
in Fig.~\ref{fig-sd-eq}.
The analytical form of this equation reads
\begin{equation}
S^{-1}(K) = S_0^{-1}(K) +4\pi\alpha_s \int\frac{d^4 p}{(2\pi)^4} 
\Gamma^{A}_{\mu} S(P) \Gamma^{B}_{\nu} {\cal D}_{AB}^{\mu\nu}(K-P),
\label{sd-eq}
\end{equation}
where the quark-gluon vertices in the Nambu-Gorkov basis are 
given by
\begin{equation}
\Gamma^{A}_{\mu} = \gamma_\mu \left(
\begin{array}{cc}
T^A & 0 \\
0 & -\left(T^A\right)^T
\end{array}
\right).
\end{equation}
In the case of the simplest ansatz for the quark propagator, as in 
Eq.~(\ref{inv_prop}), the Schwinger-Dyson equation (\ref{sd-eq}) 
reduces to an equation 
for the gap parameter $\Delta$. For this reason, it is also called 
the gap equation. After neglecting the dependence of the gap on the 
three-momentum and after performing the momentum integration, one 
arrives at the following approximate form of the gap equation in 
Euclidean space ($k_4\equiv ik_0$) \cite{weak-son,weak,weak-sp1}:
\begin{equation}
\Delta(k_4) \simeq \frac{\alpha_s}{9\pi} \int
\frac{dp_4 \Delta(p_4)}
{\sqrt{p_4^2+\Delta^2}}\ln\frac{\Lambda}{|k_4-p_4|},
\label{gap_eq-qcd}
\end{equation}
where $\Lambda = 2 (4\pi)^{3/2}\mu \alpha_s^{-5/2}$. If the quark wave 
function renormalization from the diagonal part of the self-energy were 
taken into account, the extra factor $\exp\left(-\frac{4+\pi^2}{8}\right)$ 
would appear in the expression for $\Lambda$ \cite{self-e}. 
The appearance of the logarithm in the integrand of the gap equation 
(\ref{gap_eq-qcd}) is an artifact of the long-range force in QCD, 
mediated by the magnetic gluon modes. An approximate solution to the 
gap equation reads
\begin{equation}
\Delta(0) \simeq \Lambda \exp\left(-\frac{3\pi^{3/2}}{2^{3/2}
\sqrt{\alpha_s}}\right).
\label{gap-sol}
\end{equation}
The result is plotted in Fig.~\ref{fig-gap-sol}. Here, the value of the 
coupling constant $\alpha_s\equiv \alpha_s(\mu)$ is taken at the scale 
of the quark chemical potential. 

It appears that, at densities that exist inside stars, i.e., 
$n\lesssim 10n_0$, the value of the quark chemical potential cannot 
be much larger than $500\,\mbox{MeV}$. The corresponding value 
of the QCD coupling constant $\alpha_s\equiv \alpha_s(\mu)$ is not 
small, and, in contrast to the situation in asymptotically dense QCD, 
the Cooper pairing dynamics is not weakly interacting. This suggests
that the above analysis of the Schwinger-Dyson equation may not be 
suitable for the
description of matter inside stars. In spite of this, the solution
in Eq.~(\ref{gap-sol}) is of fundamental importance in theory. This
is one of very few non-perturbative solutions in QCD which is obtained 
from first principles, and which can be systematically improved by 
studying higher order corrections. Also, this solution shows that the 
phase diagram of QCD in the plane of temperature and chemical potential 
contains color superconducting phases at least at asymptotic densities. 

 \subsubsection{Gap equation in the NJL model}

If deconfined quark matter exists in the central regions of compact 
stars, it is likely to be color superconducting even when it is not 
weakly interacting. In order to study such a strongly interacting regime
of quark matter with a typical value of the quark chemical potential
$\mu\simeq 500\,\mbox{MeV}$, one can use various effective models of 
QCD. The simplest type of such models is the NJL model with a local
four-fermion interaction in Eq.~(\ref{NJL-action}). 

The NJL model in Eq.~(\ref{NJL-action}) lacks gluons, and, as reflection 
of this, it possesses the global instead of the local (gauge) SU(3)$_{c}$
color symmetry. This model can be viewed as a result of integrating 
out heavy gluons from the QCD action. This is possible if the gluons 
get nonzero masses from non-perturbative effects. One arrives at an
effective NJL model similar to that in Eq.~(\ref{NJL-action}) when the 
gauge fixing in QCD, needed to perform the integration, is consistent 
with the global color symmetry. 

The global color symmetry of the NJL model is broken in the 2SC phase. 
Then, because of the Goldstone theorem, five massless Nambu-Goldstone 
bosons should appear in the low-energy spectrum of such a theory. In 
QCD, however, there is no room for such Nambu-Goldstone bosons. The 
seeming contradiction is removed by noting that these Nambu-Goldstone 
bosons are not physical. Their appearance 
is an artifact of the gauge fixing. In particular, there exist a gauge 
choice in QCD, namely the unitary gauge, in which these bosons can be 
completely eliminated. 

The gap equation in the NJL model in the mean field approximation 
looks as follows:
\begin{equation}
\Delta \simeq \frac{4 G_D}{\pi^2} \int_0^{\Lambda}
\left(\frac{\Delta}{\sqrt{(p-\mu)^2+\Delta^2}}
+\frac{\Delta}{\sqrt{(p+\mu)^2+\Delta^2}}\right) 
p^2 d p .
\label{gap-eq-NJL}
\end{equation}
This gap equation can be obtained from a Schwinger-Dyson equation 
similar to that in QCD in Eq.~(\ref{sd-eq}) after the gluon long-range 
interaction is replaced by a local interaction proportional to 
$\delta(K-P)$.

The approximate solution to the gap equation in Eq.~(\ref{gap-eq-NJL}) 
reads
\begin{equation}
\Delta \simeq 2\sqrt{\Lambda^2-\mu^2}
\exp\left(-\frac{\pi^2}{8G_D\mu^2}+\frac{\Lambda^2-3\mu^2}{2\mu^2}\right).
\end{equation}
This is very similar to the Bardeen-Cooper-Schrieffer solution 
in the case of low temperature superconductivity in solid state 
physics \cite{BCS}. As in the Bardeen-Cooper-Schrieffer theory, 
it has the same type 
non-analytic dependence on the coupling constant and the same 
type dependence on the density of quasiparticle states at 
the Fermi surface. 

When the quark chemical potential $\mu$ takes a value in the range 
between $400\,\mbox{MeV}$ and $500\,\mbox{MeV}$, and the strength 
of the diquark pairing is $G_D=\eta G_S$ with $\eta$ between $0.7$ and 
$1$, the value of the gap appears to be of order $100\,\mbox{MeV}$. 
In essence, this is the result that was obtained in Ref.~\cite{cs}.

 \subsubsection{Properties of quark matter in the 2SC phase}

Now let us briefly summarize the main physical properties of the 
2SC phase of dense quark matter. To large extent, these follow
directly from the symmetry of the ground state, determined by the 
structure of the condensate in Eq.~(\ref{2sc-cond}), and the 
magnitude of the diquark gap $\Delta$.

As we argued earlier, the blue up and blue down quarks do not
participate in Cooper pairing in the 2SC phase. They give rise
to gapless quasiparticles in the low energy spectrum of the theory.
The density of states of such quasiparticles is proportional to 
$\mu^2$ and, therefore, is very large. This simple fact has important 
implications for the quark matter properties. At small temperatures, 
these gapless quasiparticles give dominant contributions to the 
specific heat, as well as to the electrical and heat conductivities 
of the 2SC phase. Also, the presence of the ungapped blue up and blue 
down quarks should result in a large neutrino emissivity due to 
the $\beta$-processes, $d_b \to u_b +e^{-}+\bar{\nu}_{e}$ and
$u_b +e^{-}\to d_b +\nu_{e}$ \cite{Iwamoto}.

The other four quark quasiparticles, originating from the red and 
green paired quarks, are gapped. Their dispersion relations look
like that in Eq.~(\ref{E_qp}). At small temperatures, $T\ll \Delta$, 
the contributions of these quasiparticles to all transport and many
thermodynamic quantities are suppressed by the exponentially small 
factor $\exp\left(-\Delta/T\right)$. Gluons do not play a very 
important role either. They are bosons and their thermal number 
densities are small at small temperatures. In addition, five out 
of total eight gluons are gapped because of the color Meissner effect. 

The presence of the unpaired blue quarks is also connected with the 
absence of baryon superfluidity in the 2SC phase. The generator of 
the baryon number conservation symmetry is defined as follows: 
\begin{equation}
\tilde{B}=B-\frac{2}{\sqrt{3}}T_8 
=\mbox{diag}_{\rm color}\left(0,0,1\right),
\label{U1_B}
\end{equation}
where $B$ is the generator of the U(1)$_B$ symmetry in vacuum. This
vacuum generator $B$ mixes with the color generator $T_8$ to produce 
the generator $\tilde{B}$ of the $\tilde{\mbox{U}}(1)_B$ symmetry 
in medium. From Eq.~(\ref{U1_B}) it is clear that only (anti-)blue 
quasiparticles carry a nonzero baryon number in the 2SC phase.

In the 2SC phase, there is also an unbroken $\tilde{\mbox{U}}(1)_{\rm em}$ 
gauge symmetry. The corresponding generator reads
\begin{equation}
\tilde{Q}=Q-\frac{1}{\sqrt{3}}T_8 ,
\label{U1_em_2sc}
\end{equation}
where $Q=\mbox{diag}_{\rm flavor}(\frac{2}{3},-\frac{1}{3})$ is the 
generator of the U(1)$_{\rm em}$ symmetry in vacuum. The gauge boson 
of $\tilde{\mbox{U}}(1)_{\rm em}$ is nothing else but the medium 
photon. It should be clear, therefore, that the 2SC ground state 
is not subject to the electromagnetic Meissner effect. Thus, if 
there is an external magnetic field, it is not expelled from such 
a medium.

One should notice that the quark quasiparticles carry the following 
$\tilde{Q}$-charges:
\begin{eqnarray}
\tilde{Q}(u_r)=\tilde{Q}(u_g)=\frac{1}{2}, && \tilde{Q}(u_b)=1,\\
\tilde{Q}(d_r)=\tilde{Q}(d_g)=-\frac{1}{2}, && \tilde{Q}(d_b)=0.
\end{eqnarray}
It is interesting that the blue up quark, which remains unpaired 
in the 2SC phase, is charged (this is not true for the blue down
quark which is also unpaired). It means that this phase of matter
is a $\tilde{Q}$-conductor, and that the value of the electrical 
conductivity is very high. 

The pressure of cold dense quark matter is dominated by the Pauli 
pressure. The partial contribution of each quark is $\mu^4/12\pi^2$,
see Eq.~(\ref{pressure_0}). The correction due to color superconductivity 
is parametrically suppressed by the factor $\left(\Delta/\mu\right)^2$. 
The reason is that the Cooper pairing affects only those quark states 
that are close to the Fermi surface. In contrast, it is the whole Fermi 
sphere that contributes to the Pauli pressure. At weak coupling, 
the contribution due to diquark pairing can be easily calculated. 
It is given by $\left(\mu\Delta/2\pi\right)^2$ per each gapped quark 
quasiparticle \cite{eff-pot}. In the 2SC phase, there are six (two 
flavors times three colors) quarks in total, and four of them give 
rise to quasiparticles with the gap $\Delta$ in their energy spectra. 
Thus, the pressure of such matter is approximately given by 
\cite{eff-pot,kappa}
\begin{equation}
P_{\rm (2SC)}\simeq \frac{N_c N_f \mu^4}{12\pi^2} -B 
+ 4 \left(\frac{\mu\Delta}{2\pi}\right)^{2}
=\frac{\mu^4}{2\pi^2} -B 
+ \frac{\mu^2\Delta^2}{\pi^2}.
\label{pres-2sc}
\end{equation}
Note that the bag pressure was also added on the right hand side. If 
present, this term partially cancels the leading $\mu^4$ term. Because 
of this, color superconductivity may have a large effect on the 
hadron-quark phase transition \cite{LH02} and even on the properties 
of compact stars \cite{AlRe03}. By making use of thermodynamic 
identities, one can derive the corresponding energy density,
\begin{equation}
\epsilon_{\rm (2SC)}\simeq \frac{3\mu^4}{2\pi^2} +B
+ \frac{\mu^2\Delta^2}{\pi^2}
\left(1+\frac{2\mu}{\Delta}\frac{\partial\Delta }{\partial\mu}\right).
\label{ener-2sc}
\end{equation}
The expressions in Eqs.~(\ref{pres-2sc}) and (\ref{ener-2sc}) give a 
parametric representation of the equation of state of dense quark 
matter in the 2SC phase. Such an equation of state is the key input
in the Tolman-Oppenheimer-Volkoff equations \cite{Tol} which
determine the interior structure of compact stars. In short, such 
equations describe hydrostatic equilibrium of matter inside a star 
when the pressure gradient in each radial layer of the star is 
balanced by the gravitational weight of the layer itself. Several 
constructions of compact stars with quark matter in their interior 
were presented in Refs.~\cite{LH02,AlRe03,BB02,BBBNOS02,SHH,BGAYT,BNOS}.

\subsection{Color-flavor locked phase ($N_f=3$)}

It may happen that dense baryonic matter is made not only of 
the lightest up and down quarks, but of strange quarks as well.
In fact, because of a possible reduction in the free energy 
from converting non-strange quarks into strange quarks, one may 
even speculate that strange quark matter is the true ground state of 
baryonic matter \cite{Bod}. 

The constituent strange quark mass in vacuum QCD is estimated to be 
of order $500\,\mbox{MeV}$. Its current mass is about $100\,\mbox{MeV}$. 
In dense baryonic matter (say, with $\mu\simeq 500\,\mbox{MeV}$), the 
value of the strange quark mass should be in the range between 
$100\,\mbox{MeV}$ and $500\,\mbox{MeV}$. It is plausible then that 
strange quarks also participate in Cooper pairing.

Here, we consider an idealized version of three-flavor quark matter,
in which all quarks are assumed to be massless. The more realistic 
case of a nonzero strange quark mass will be briefly discussed 
in the second lecture. In the massless case, the quark model 
possesses the global SU(3)$_L\times$SU(3)$_R$ chiral symmetry and the 
global U(1)$_B$ symmetry connected with the baryon number conservation.
This is in addition to SU(3)$_c$ color gauge symmetry. Note that the 
generator $Q=\mbox{diag}_{\rm flavor}(\frac{2}{3},-\frac{1}{3},
-\frac{1}{3})$ of the U(1)$_{\rm em}$ symmetry of electromagnetism 
is traceless, and therefore it coincides with one of the vector-like
generators of the SU(3)$_L\times$SU(3)$_R$ chiral group.

To large extent, the color and flavor structure of the diquark 
condensate of Cooper pairs in the three-flavor quark matter is
again determined by the symmetry of the attractive diquark channel
(i.e., color-antisymmetric antitriplet) and the Pauli exclusion 
principle. The spin-0 condensate corresponds to the following 
ground state expectation value \cite{cfl}:
\begin{equation}
\left\langle \left(\bar{\psi}^C\right)_i^a \gamma^5 \psi_j^b 
\right\rangle
\sim \sum_{I,J=1}^{3} c^{I}_{J}\varepsilon_{ijI} \epsilon^{abJ} 
+\cdots,
\label{cfl-cond}
\end{equation}
which is antisymmetric in the color and flavor indices of the 
constituent quarks, cf. Eq.~(\ref{2sc-cond}). The $3\times 3$
matrix $c^{I}_{J}$ is determined from the global minimum of the free 
energy. It appears that $c^{I}_{J}=\delta^{I}_{J}$. The ellipsis 
on the right hand side stand for a contribution which is symmetric 
in color and flavor. A small contribution of this type is always 
induced in the ground state, despite the fact that it corresponds 
to a repulsive diquark channel \cite{cfl,weak-cfl}. This is not 
surprising after noting that the symmetric condensate $\left\langle 
\left(\bar{\psi}^C\right)_i^a \gamma^5 \psi_j^b \right\rangle
\sim \delta_{i}^{a}\delta_{j}^{b}+\delta_{j}^{a}\delta_{i}^{b}$
does not break any additional symmetries \cite{cfl}. 

In the ground state, determined by the condensate (\ref{cfl-cond}), 
the chiral symmetry is broken down to its vector-like subgroup.
However, the mechanism of this symmetry breaking is very unusual. 
To see this, let us rewrite the condensate in Eq.~(\ref{cfl-cond})
as follows:
\begin{equation}
\left\langle \psi_{L,i}^{a,\alpha} \epsilon_{\alpha\beta}
\psi_{L,j}^{b,\beta}\right\rangle
=-\left\langle \psi_{R,i}^{a,\dot\alpha} \epsilon_{\dot\alpha\dot\beta}
\psi_{R,j}^{b,\dot\beta}\right\rangle
\sim \sum_{I=1}^{3} \varepsilon_{ijI} \epsilon^{abI} 
+\cdots,
\label{LL-RR-cond}
\end{equation}
where $\alpha,\beta,\dot\alpha,\dot\beta=1,2$ are the spinor indices. 
The condensate of left-handed fields in Eq.~(\ref{LL-RR-cond}) breaks 
the SU(3)$_c$ color symmetry and the SU(3)$_L$ chiral symmetry, but 
leaves the diagonal SU(3)$_{L+c}$ subgroup unbroken. Indeed, as one 
can check, this condensate remains invariant under the simultaneous 
flavor transformation $g_L$ and the compensating color transformation 
$g_c=(g_L)^{-1}$. Similarly, the condensate of right-handed fields in 
Eq.~(\ref{LL-RR-cond}) leaves the SU(3)$_{R+c}$ subgroup unbroken. 

When both condensates are present,
the symmetry of the ground state is SU(3)$_{L+R+c}$. At the level of 
global symmetries, the original SU(3)$_{L}\times$SU(3)$_{R}$ is broken
down to the vector-like SU(3)$_{L+R}$, just like in vacuum. Unlike 
in vacuum, however, this breaking does {\em not} result from any 
condensates mixing left- and right-handed fields. Instead, it results 
from two separate condensates, made of left-handed fields only and 
of right-handed fields only. The color-flavor orientations of the two 
condensates are ``locked'' to each other by color transformations. 
This mechanism is called locking, and the corresponding phase of 
matter is called color-flavor-locked (CFL) phase \cite{cfl}.

To large extent, the gap equation in the three-flavor quark matter
is the same as in the two-flavor case. The differences come only 
from a slightly more complicated color-flavor structure of the 
off-diagonal part of the inverse quark propagator (gap matrix) 
\cite{cfl,weak-cfl},
\begin{equation}
\Delta^{ij}_{ab}=i\gamma^5\left[
\frac{1}{3}\left(\Delta_1+\Delta_2\right)\delta^{i}_{a}\delta^{j}_{b}
-\Delta_2\delta^{i}_{b}\delta^{j}_{a}\right],
\label{delta_cfl}
\end{equation}
where two parameters $\Delta_1$ and $\Delta_2$ determine the values 
of the gaps in the quasiparticles spectra. In the ground state that 
respects the SU(3)$_{L+R+c}$ symmetry, the original nine quark 
states give rise to a singlet and an octet of quasiparticles. The 
singlet has gap $\Delta_1$, and the octet has gap $\Delta_2$.

When the small symmetric diquark condensate is neglected, one finds
that $\Delta_1=2\Delta_2$, i.e., the gap of the singlet is twice as 
large as the gap of the octet. In real case, however, this relation
holds only approximately. The value of the octet gap $\Delta_2$ in 
asymptotically dense QCD is given by an expression similar to that 
in Eq.~(\ref{gap-sol}), but the overall coefficient is multiplied 
by an extra factor $2^{-1/3}\left(2/3\right)^{5/2}\approx 0.288$ 
\cite{weak-cfl,loops}.

 \subsubsection{Properties of quark matter in the CFL phase}

Let us briefly review physical properties of the CFL phase. In 
contrast to the 2SC phase, there are no gapless quark quasiparticles 
in the low energy spectrum of the CFL phase. This implies that, 
at small temperatures, $T\ll \Delta$, the contributions of quark 
quasiparticles to all transport and many thermodynamic quantities 
are suppressed by the exponentially small factor 
$\exp\left(-\Delta/T\right)$. Gluons do not play any important 
role either. All of them are gapped because of the color Meissner 
effect. 

Unlike the 2SC phase, the CFL phase is superfluid. This is because 
the U(1)$_B$ baryon number symmetry is broken in the ground state. 
If such a phase appears in a core of a rotating star, it will be 
threaded with rotational vortices. This may be related to the 
existence of such phenomena as glitches. These are sudden changes 
of the rotational frequency observed in some pulsars. The glitches 
may be caused by occasional releasing of the angular momentum of 
rotational vortices, that remained pinned to the stellar crust 
\cite{gl-super}.

Because of the Goldstone theorem, the breaking of the U(1)$_B$ 
baryon number symmetry should result in the appearance of a 
Nambu-Goldstone boson in the low-energy theory. In absence of 
gapless quark quasiparticles, this Nambu-Goldstone boson turns 
out to play an important role in many transport properties of cold 
CFL matter \cite{kappa,opaque}.

The CFL phase, like the 2SC phase, is not an electromagnetic 
superconductor. Therefore, it does not expel a magnetic flux
from its interior. This is the result of having an unbroken 
$\tilde{\mbox{U}}(1)_{\rm em}$ gauge symmetry in the ground 
state. The corresponding generator reads
\begin{equation}
\tilde{Q}=Q-T_3-\frac{1}{\sqrt{3}}T_8 .
\label{U1_em_cfl}
\end{equation}
The quark quasiparticles carry the following $\tilde{Q}$-charges:
\begin{eqnarray}
\tilde{Q}(u_r)=0, && \tilde{Q}(u_g)=\tilde{Q}(u_b)=1,\\
\tilde{Q}(d_r)=-1, && \tilde{Q}(d_g)=\tilde{Q}(d_b)=0,\\
\tilde{Q}(s_r)=-1, && \tilde{Q}(s_g)=\tilde{Q}(s_b)=0.
\end{eqnarray}
Since all quasiparticles are gapped, the corresponding phase of 
matter is a $\tilde{Q}$-insulator. Its electrical conductivity at 
small temperatures is dominated by thermally excited electrons and 
positrons \cite{opaque}. At sufficiently low temperatures, when the 
thermal density is low, the CFL phase becomes transparent to light
\cite{opaque,opaque2}. 

In order to write down the expression for the pressure of three-flavor 
quark matter in the CFL phase, we take the Pauli pressure contributions 
of nine (three flavors times three colors) quarks and add the correction 
due to color superconductivity from one quasiparticle with gap $\Delta_1$ 
and eight quasiparticles with gap $\Delta_2$. Thus, we derive
\begin{eqnarray}
P_{\rm (CFL)} &\simeq &\frac{N_c N_f \mu^4}{12\pi^2} -B 
+ \left(\frac{\mu\Delta_1}{2\pi}\right)^{2}
+ 8\left(\frac{\mu\Delta_2}{2\pi}\right)^{2}
\nonumber \\
&\simeq& \frac{3\mu^4}{4\pi^2} -B + 3\frac{\mu^2\Delta^2}{\pi^2},
\label{pres-cfl}
\end{eqnarray}
where we used the approximate relation between the singlet and 
octet gaps, $\Delta_1=2\Delta_2\equiv 2\Delta$.
By making use of thermodynamic identities, one can derive the 
corresponding energy density,
\begin{equation}
\epsilon_{\rm (CFL)}\simeq \frac{9\mu^4}{4\pi^2} +B
+ 3\frac{\mu^2\Delta^2}{\pi^2}
\left(1+\frac{2\mu}{\Delta}\frac{\partial\Delta }{\partial\mu}\right).
\label{ener-cfl}
\end{equation}
These two expressions give a parametric representation of the 
equation of state of dense quark matter in the CFL phase. 

 \subsubsection{Low-energy effective action}

As in the case of chiral perturbation theory in vacuum QCD \cite{GasLeut}, 
one could write down an effective low-energy theory in the CFL phase. 
From the symmetry breaking pattern, it is known that there are nine 
Nambu-Goldstone bosons and one pseudo-Nambu-Goldstone boson in the 
low-energy spectrum of the theory \cite{HRZ,CasGat}. Eight of the 
Nambu-Goldstone bosons are similar to those in vacuum QCD: three pions
($\pi^0$ and $\pi^\pm$), four kaons ($K^0$, $\bar{K}^0$ and $K^\pm$)
and the eta-meson ($\eta$). The additional Nambu-Goldstone boson ($\phi$) 
comes from breaking the U(1)$_B$ baryon symmetry. Finally, the 
pseudo-Nambu-Goldstone boson ($\eta^\prime$) results from breaking 
the approximate axial U(1)$_A$ symmetry.

The low energy action for the Nambu-Goldstone bosons in the CFL phase was 
derived in Refs.~\cite{HRZ,CasGat,SonSt}. In the chiral limit, the 
result reads
\begin{eqnarray}
{\cal L}_{eff} &=& \frac{f_\pi^2}{4} {\rm Tr}\left[
 \partial_0\Sigma\partial_0\Sigma^\dagger - v^2
 \partial_i\Sigma\partial_i\Sigma^\dagger \right]
+ \frac{1}{2} \left[ (\partial_0 \phi)^{2}
- v^2 (\partial_i\phi)^{2} \right]
\nonumber \\
&+&\frac{1}{2} \left[ (\partial_0 \eta^{\prime})^{2}
- v^2 (\partial_i\eta^{\prime})^{2} \right],
\label{cov-der}
\end{eqnarray}
where $f^2_\pi=(21-8\ln 2)(\mu/6\pi)^2$ and $v^2=1/3$ were 
calculated in asymptotically dense QCD in Refs.~\cite{SonSt,fpi-more}. 
(A nonzero mass of the $\eta^\prime$-meson was neglected here 
\cite{SSZh}.) By definition, 
$\Sigma\equiv \exp \left(i\lambda^A \pi^A/f_\pi\right)$ 
is a unitary matrix field which describes the octet of the 
Nambu-Goldstone bosons, transforming under the chiral 
$SU(3)_{L} \times SU(3)_{R}$ group as follows:
\begin{equation}
\Sigma \to U_{L} \Sigma U_{R}^{\dagger},
\end{equation}
where $(U_{L},U_{R}) \in SU(3)_{L} \times SU(3)_{R}$. 

The extra terms in the low-energy action should be added when 
quark masses are nonzero \cite{SonSt,mmm}. Since the quark 
masses break the chiral symmetry of dense QCD explicitly, these type
of corrections produce non-vanishing masses for all Nambu-Goldstone 
bosons, except one. The single boson left massless is the 
Nambu-Goldstone boson related to the baryon number breaking. 
Of course, this symmetry is not affected by the quark masses.

Within the framework of the high density effective theory 
\cite{loops,hdet}, it was shown that an additional effect 
of nonzero quark masses may appear when the strange 
quark mass exceeds a critical value $m_s\sim m_u^{1/3}\Delta^{2/3}$. 
In this case, the CFL phase is expected to undergo a phase transition 
to a phase with a meson (e.g., kaon or/and eta) condensate 
\cite{BS,KR,Kry2004cw}. In the low-energy action of the corresponding 
phase, additional Nambu-Goldstone bosons with very unusual properties 
can appear \cite{abnormal}.

\subsection{Spin-1 color superconductivity ($N_f=1$)}

In the case of neutral matter in $\beta$ equilibrium, as we shall discuss 
in the second lecture in more detail, it may happen that different
quark flavors cannot create Cooper pairs because of a large mismatch 
between their Fermi momenta. In this case, one could consider the
possibility of a much weaker spin-1 Cooper pairing 
\cite{Bail,weak-sp1,sp1995,spin-1,spin-1-Meissner,andreas}. It is the Pauli 
principle that does not allow to construct spin-0 Cooper pairs from 
quarks of the same flavor. The color antisymmetric wave function of 
a pair can only be symmetric in spin indices. This corresponds to
a spin-1 state.

The pairing in a spin-1 triplet state and in a color antisymmetric 
antitriplet state can lead to a rather complicated structure of the
diquark condensate. The general structure of the gap matrix can be 
written in the following form \cite{andreas}:
\begin{equation}
\Delta^{ab}=i\Delta_0 \sum_{c,i=1}^{3} \epsilon^{abc} {\cal C}_{ci} 
\left[\hat{k}^i\cos\theta +\gamma^i_\perp\sin\theta\right],
\label{delta_1sc}
\end{equation}
where $\hat{\mathbf{k}}\equiv \mathbf{k}/k$, $\gamma^i_\perp\equiv 
\gamma^i-\hat{k}^i (\mbox{\boldmath $\gamma$}\cdot\hat{\mathbf{k}})$, and the 
explicit form of the $3\times 3$ matrix ${\cal C}_{ci}$ as well as
the angular parameter $\theta$ are determined by the minimization 
of the quark matter free energy. In the special cases with $\theta=0$
and $\theta=\pi/2$, the corresponding gaps are called longitudinal and 
transverse, respectively.

By choosing various specific matrices ${\cal C}$, one could construct 
many different spin-1 phases. The four most popular of them are the 
so-called A-phase, the color-spin-locked (CSL) phase, the polar and 
the planar phases. These are determined by the following matrices
\cite{andreas}:
\begin{eqnarray}
{\cal C}^{\rm (A-phase)} 
=\frac{1}{\sqrt{2}}\left(\begin{array}{lll}
0 & 0 & 0 \\
0 & 0 & 0 \\
1 & i & 0 
\end{array}\right), &\quad &
{\cal C}^{\rm (CSL)} 
=\frac{1}{\sqrt{3}}\left(\begin{array}{lll}
1 & 0 & 0 \\
0 & 1 & 0 \\
0 & 0 & 1 
\end{array}\right), \\
{\cal C}^{\rm (polar)} 
=\left(\begin{array}{lll}
0 & 0 & 0 \\
0 & 0 & 0 \\
0 & 0 & 1 
\end{array}\right),&\quad &
{\cal C}^{\rm (planar)} 
=\frac{1}{\sqrt{2}}\left(\begin{array}{lll}
1 & 0 & 0 \\
0 & 1 & 0 \\
0 & 0 & 0 
\end{array}\right).
\end{eqnarray}
The corresponding four phases are characterized by different symmetries 
of their ground state. In particular, the original group SU(3)$_c\times 
$SO(3)$_J\times $U(1)$_{\rm em}$ of one-flavor quark matter breaks down 
as follows \cite{spin-1,spin-1-Meissner,andreas}: 
\begin{eqnarray}
\mbox{A-phase:} & &
SU(2)_c\times \widetilde{SO}(2)_J \times \tilde{U}(1)_{\rm em},\\
\mbox{CSL:} & & \widetilde{SO}(3)_{J},\label{csl}\\
\mbox{Polar:} & &
SU(2)_c\times SO(2)_J \times \tilde{U}(1)_{\rm em},\\
\mbox{Planar:} & &
\widetilde{SO}(2)_J \times \tilde{U}(1)_{\rm em}.
\end{eqnarray}
It was argued recently that the phase with the lowest free energy 
is the transverse A-phase \cite{andreas}. 

The value of the spin-1 gap is estimated to be about two or three orders 
of magnitude smaller than a typical spin-0 gap. At best, it can be about 
$1\,\mbox{MeV}$, but more realistically it is expected to be about
$0.1\,\mbox{MeV}$, or less. The presence of such a small gap is unlikely 
to modify considerably any thermodynamic, or even transport properties 
of dense quark matter. It is fair to mention, however, that the cooling 
history of stars might be sensitive enough to feel the effect of such 
small gaps \cite{B_cool}. It was also suggested that the electromagnetic 
properties of spin-1 phases can be of phenomenological importance 
\cite{spin-1-Meissner}. 
In contrast to the 2SC and CFL phases, discussed earlier, spin-1 
phases reveal the electromagnetic Meissner effect. As one can see from 
Eq.~(\ref{csl}), this clearly applies to the CSL phase. It turns out that 
the electromagnetism is also subject to the Meissner effect when there 
appears a mixture of any two independent spin-1 condensates made of 
quarks with different charges (e.g., one is made of up quarks and the 
other is made of down quarks) \cite{spin-1-Meissner}. The type-I 
superconductivity in such a system could have observable effects on 
the magnetic field relaxation in pulsars.

\subsection{Summary of the first lecture}

In this lecture, only the main three color superconducting phases of 
dense quark matter have been discussed. These are the simplest possible 
phases that can be realized in one-, two-, and three-flavor quark matter. 
It should be emphasized, however, that the conditions in dense matter 
were idealized by assuming that the chemical potentials of pairing 
quarks were equal. In real 
situation which, for example, is realized in the interior of compact 
stars, matter happens to be very different from ideal. In particular, 
the charge neutrality and the $\beta$ equilibrium are two very important 
conditions that may modify the quark chemical potentials and, therefore,
affect the properties of quark matter. These issues are discussed in 
the second lecture.

The aim of this lecture was to introduce the general idea of color 
superconductivity in cold dense baryonic matter. It was argued that 
such matter is expected to be deconfined and, because of the Cooper 
instability, it should be color superconducting. To large extent, 
this phenomenon is the same as low-temperature superconductivity 
in condensed matter physics \cite{BCS}. One feature in dense 
QCD is very remarkable. The attractive interaction between quarks 
comes from the gauge boson exchange. This is in contrast to the 
Bardeen-Cooper-Schrieffer
theory in solid state physics, where the gauge boson (photon) exchange 
is repulsive, while an attractive interaction comes from the phonon 
exchange. 

One of the key points of this lecture is the observation that QCD at
asymptotic densities is a weakly interacting (although non-perturbative) 
theory. Moreover, the properties of its color superconducting ground state 
can be studied analytically from first principles, providing a rare 
example of an essentially solvable limit in a non-Abelian theory. By 
itself, this has a fundamental theoretical importance. Also, this result 
may provide valuable insights in the theory of strong interactions. 
One of the examples might be the idea of duality between the hadronic 
and quark description of QCD \cite{cont}.

Although the use of weakly interacting QCD for the description of dense
quark matter is very instructive by itself, it does not seem to be very 
useful for the quantitative description of the ground state at densities 
existing inside stars. The reason is that the density of matter in stars 
is bounded from above by the condition of the hydrostatic equilibrium, 
$n\lesssim 10n_0$. At such not so large densities, one cannot treat QCD 
as a weakly interacting theory. In this situation, the use of various 
effective models of QCD has proved to be very useful.

\newpage 
\section{Color superconductivity in neutral matter}

\subsection{Dense matter inside stars}

As we discussed in the first lecture, it is natural to expect that 
some color superconducting phases may exist in the interior of compact 
stars. The estimated central densities of such stars might be sufficiently
large for producing deconfined quark matter. Such matter should develop
the Cooper instability and become color superconducting. It should
also be noted that typical temperatures inside compact stars are so 
low that the diquark condensate, if produced, would not melt.

In the first lecture, we discussed the idealized version of dense matter, 
in which the Fermi momenta of pairing quarks were assumed to be equal.
This does not describe the real situation that should occur inside 
compact stars. The reason is that matter in the bulk of a compact star 
should be neutral (at least, on average) with respect to electric as 
well as color charges. Otherwise, the star would not be bound by gravity 
which is much weaker than electromagnetism. Matter should also remain 
in $\beta$ equilibrium, i.e., all $\beta$ processes (e.g., such as 
$d \to u + e^{-} + \bar\nu_{e}$,     $u + e^{-} \to d + \nu_{e}$, 
$s \to u + e^{-} + \bar\nu_{e}$, and $u + e^{-} \to s + \nu_{e}$) 
should go with equal rates in both directions. 

As we shall see below, after the charge neutrality and the $\beta$ 
equilibrium in quark matter are enforced, the chemical potentials of 
different quarks would satisfy some relations that may interfere 
with the dynamics of Cooper pairing. If this happens, some color 
superconducting phases may become less favored than others. For 
example, it was argued in Ref.~\cite{no2sc}, that a mixture 
of unpaired strange quarks and the non-strange 2SC phase, made of 
up and down quarks, is less favorable than the CFL phase after the 
charge neutrality condition is enforced. A similar conclusion was 
also reached in Ref.~\cite{n_steiner}.

Assuming that the constituent medium modified mass of the strange
quark is large (i.e., larger than the corresponding strange quark
chemical potential), in Ref.~\cite{SH} it was shown that neutral 
two-flavor quark matter in $\beta$-equilibrium can have another 
rather unusual ground state called the gapless two-flavor color 
superconductor (g2SC). While the symmetry in the g2SC ground state 
is the same as that in the conventional 2SC phase, the spectrum of 
the fermionic quasiparticles is different.

The existence of gapless color superconducting phases was confirmed
in Refs.~\cite{GLW,var-appr,rusterR}, and generalized to nonzero temperatures
in Refs.~\cite{HS,LZ}. In addition, it was also shown that a gapless
CFL (gCFL) phase could appear in neutral strange quark matter
\cite{gCFL,gCFL-long}. At nonzero temperature, the gCFL phase
and several other new phases (e.g., the so-called dSC and uSC phases)
were studied in Refs.~\cite{dSC,RSR,FKR}. If the surface tension is
sufficiently small, as suggested in Ref.~\cite{RR}, the mixed phase
composed of the 2SC phase and the normal quark phase \cite{SHH} may 
be more favored. 

Here, it may be appropriate to mention that a non-relativistic
analogue of gapless superconducting phases could appear in a trapped
gas of cold fermionic atoms \cite{WilLiu,Deb,LWZ,FGLW,GWM}. (Note that an
alternative ground state for the atomic system, similar to the quark
mixed phases in Refs.~\cite{RR,SHH,mixed}, was proposed in
Ref.~\cite{Bed}.)

\subsection{Gapless two-flavor color superconductivity}

 \subsubsection{Neutrality vs. color superconductivity}

Let us discuss the effect of charge neutrality on color superconductivity.
It is instructive to start with the case of two-flavor quark matter first. 
In this case, the effect is most prominent and, thus, it is easiest to 
understand. Later, the results for three-flavor quark matter will be also 
discussed.

In order to get an impression about the importance of the charge
neutrality condition in a large macroscopic chunk of matter, such as 
a core of a compact star, let us estimate the corresponding Coulomb 
energy. A simple calculation leads to the following result:
\begin{equation}
E_{\rm Coulomb} \sim n_Q^2 R^5 \sim M_\odot c^2
\left(\frac{n_Q}{10^{-15} e/\mbox{fm}^3}\right)^2
\left(\frac{R}{1\,\mbox{km}}\right)^5,
\label{E_Coulomb}
\end{equation}
where $R$ is the radius of the quark matter core, whose charge density
is denoted by $n_Q$. It is easy to see that this energy is not an 
extensive quantity: the value of the corresponding {\em energy density}
increases with the size of the system as $R^2$. By taking a typical 
value of the charge density in the ideal 2SC phase, $n_Q\sim 10^{-2} 
e/\mbox{fm}^3$, the energy in Eq.~(\ref{E_Coulomb}) becomes a factor 
of $10^{26}$ larger than the rest mass energy of the Sun\,! To avoid 
such an incredibly large energy price, the charge neutrality $n_Q=0$ 
should be satisfied with a very high precision. 

In the case of two-flavor quark matter, one can argue that the neutrality 
is achieved if the number density of down quarks is approximately 
twice as large as number density of up quarks, $n_d \approx 2n_u$. 
This follows from the fact that the negative charge of the down quark 
($Q_d=-1/3$) is twice as small as the positive charge of the up quark 
($Q_u=2/3$). When $n_d \approx 2n_u$, the total electric charge 
density is vanishing in absence of electrons, $n_Q\approx Q_d n_d
+Q_u n_u\approx 0$. It turns out that even a nonzero density of electrons, 
required by the $\beta$ equilibrium condition, could not change this 
relation much. 

The argument goes as follows. Let us consider noninteracting massless 
quarks. In $\beta$ equilibrium, the chemical potentials of the up quark 
and the down quark, $\mu_u$ and $\mu_d$, should satisfy the relation 
$\mu_d=\mu_u+\mu_e$ where $\mu_e$ is the chemical potential of 
electrons (i.e., up to a sign, the chemical potential of the 
electric charge). By assuming that $\mu_d\approx 2^{1/3}\mu_u$, i.e., 
$n_d\approx2n_u$ as required by the neutrality in absence of electrons, 
one obtains the following result for the electron chemical potential:
$\mu_e=\mu_d-\mu_u \approx \frac{1}{4}\mu_u$. The corresponding 
density of electrons is $n_e\approx 5\cdot 10^{-3}n_u$. We see that
$n_e\ll n_u$ which is in agreement with the original assumption that 
$n_d\approx 2n_u$ in neutral matter. 

While the approximate relation $n_d\approx 2n_u$ may be slightly 
modified in an interacting system, the main conclusion would remain
qualitatively the same. The Fermi momenta of up and down quarks, whose 
pairing is responsible for color superconductivity, are generally 
non-equal after the neutrality and $\beta$ equilibrium conditions 
are imposed. In this case, the Cooper pairing may be substantially 
modified, or even prevented. The study of this issue is the topic 
of the rest of this lecture.

\subsubsection{Model}

In view of strongly coupled dynamics and various non-perturbative 
(e.g., instanton) effects in dense QCD at moderate densities
$n\lesssim 10n_0$, existing in central regions of compact stars, 
there is no real advantage in using microscopic theory of QCD. 
For purposes of the current presentation, it suffices to use an 
effective model such, for example, as the NJL model with the 
Lagrangian density in Eq.~(\ref{NJL-action}), in which the common
quark chemical potential $\mu$ is replaced by a properly chosen 
color-flavor matrix $\hat\mu$ of chemical potentials 
\cite{no2sc,n_steiner,huang_2sc}.

In $\beta$ (chemical) equilibrium, the matrix $\hat\mu$ is determined
only by a few independent chemical potentials which can be introduced 
in the corresponding partition function as Lagrange multipliers in 
front of conserved charges $Q_i$ of the model,
\begin{equation}
Z=\mbox{Tr}\,\exp\left(-\frac{H+\sum_i \mu_{i} Q_i}{T}\right),
\end{equation}
In two-flavor quark matter, 
we consider only three relevant conserved charges: the baryon number 
$n_B$, the electric charge $n_Q$, and the color charge $n_{8}$. Then, 
the matrix of quark chemical potentials is given in terms of the baryon 
chemical potential (by definition, $\mu_B\equiv 3\mu$), the electron 
chemical potential ($\mu_e$) and the color chemical potential ($\mu_8$)
\cite{SH},
\begin{equation}
\hat\mu_{ij, \alpha\beta}= (\mu \delta_{ij}- \mu_e Q_{ij})
\delta_{\alpha\beta} + \frac{2}{\sqrt{3}}\mu_{8} \delta_{ij}
(T_{8})_{\alpha \beta},
\end{equation}
where $Q$ and $T_8$ are the generators of U(1)$_{\rm em}$ of
electromagnetism and the U(1)$_{8}$ subgroup of SU(3)$_{c}$ color 
gauge group. The explicit expressions for chemical potentials of 
different quarks read
\begin{eqnarray}
\mu_{ur} =\mu_{ug} =\mu -\frac{2}{3}\mu_{e} +\frac{1}{3}\mu_{8},
\label{mu-ur-ug} \\
\mu_{dr} =\mu_{dg} =\mu +\frac{1}{3}\mu_{e} +\frac{1}{3}\mu_{8}, 
\label{mu-dr-dg} \\
\mu_{ub} =\mu -\frac{2}{3}\mu_{e} -\frac{2}{3}\mu_{8}, 
\label{mu-ub} \\
\mu_{db} =\mu +\frac{1}{3}\mu_{e} -\frac{2}{3}\mu_{8}.
\label{mu-db} 
\end{eqnarray}
Note that there exist two mutually commuting color charges in the 
model. They are related to the generators $T_3$ and $T_8$ of the 
SU(3)$_{c}$ gauge group. Then, generally speaking, one could 
introduce two independent chemical potentials for the corresponding 
two color charges. Because of the invariance of the 2SC ground state 
under the transformations of the SU(2)$_{c}$ color gauge subgroup, 
the introduction of the second nontrivial color chemical potential 
$\mu_{3}$ is not necessary. This additional color chemical potential, 
however, is generally needed in three-flavor quark matter.

Here, several comments are in order regarding the introduction of 
color quark chemical potential $\mu_8$. This is required for enforcing 
color charge neutrality only in the color superconducting ground state. 
In the NJL model, it is added by hand in the partition function, and 
then adjusted to neutralize the ground state. In QCD, on the other 
hand, the charge neutrality is realized dynamically due to the generation 
of the gluon condensate $\langle A_0^8\rangle\neq 0$ \cite{DD}. 
The appearance of such a condensate is equivalent to having 
a nonzero value of the chemical potential $\mu_8\sim g_s \langle 
A_0^8\rangle$. A similar relation exists between the chemical 
potential $\mu_3$ and the gluon condensate $\langle A_0^3\rangle$ 
in three-flavor quark matter. It should be clear, therefore, that 
nonzero gluon condensates, such as $\langle A_0^8\rangle$ or 
$\langle A_0^3\rangle$, do not break explicitly any symmetries in 
QCD. They appear spontaneously only in the color superconducting 
ground state, and only after gauge fixing.

\subsubsection{Effective potential}

In order to derive the effective potential in the NJL model 
(\ref{NJL-action}) in the mean field approximation, it is convenient 
to use the standard Hubbard-Stratonovich method \cite{Hub-Stra}. Then, 
the original NJL Lagrangian density is replaced by
\begin{eqnarray}
\tilde{\cal L}_{\rm NJL} &=& \bar\psi\left[i\gamma^\mu \partial_\mu
+ \gamma^0 \hat{\mu} - m^{(0)} -\sigma
-i\gamma_5 (\vec{\mbox{\boldmath $\pi$}} \cdot\vec{\tau})\right]\psi 
-\frac{\sigma^2+\vec{\mbox{\boldmath $\pi$}}^2}{4G_S}
\nonumber\\
&-& \frac{\phi_a^{*} \phi^a}{4G_D}
-\frac{i}{2} \bar{\psi}^C \gamma_5 \varepsilon \epsilon^{a}
\phi^a \psi
-\frac{i}{2} \bar{\psi} \varepsilon \epsilon^{a} \phi_a^{*}
\gamma_5 \psi^C ,
\label{HS-action}
\end{eqnarray}
where $\sigma$, $\vec{\mbox{\boldmath $\pi$}}$, and $\phi_a$ are 
auxiliary fields. After taking into account the equations of motion 
for these fields,
\begin{eqnarray}
\sigma &=& - 2 G_S \left(\bar\psi \psi\right) ,\\
\vec{\mbox{\boldmath $\pi$}} &=& - 2 G_S\left(\bar\psi i\gamma_5 
\vec{\tau} \psi\right),\\
\phi^a &=& - 2G_D \left( i \bar{\psi} \varepsilon
\epsilon^{a} \gamma_5 \psi^C \right) ,\\
\phi_a^{*} &=& - 2G_D \left(i \bar{\psi}^C \gamma_5 \varepsilon 
\epsilon^{a} \psi\right) ,
\end{eqnarray}
one can check that the Lagrangian density in (\ref{HS-action}) is 
equivalent to the original NJL model. In the mean field approximation,
the auxiliary fields are replaced by their vacuum expectation values,
$\langle\sigma\rangle = m-m^{(0)}$, 
$ \langle \vec{\mbox{\boldmath $\pi$}} \rangle=0$ and 
$\langle \phi^a \rangle=\Delta\delta^{a}_{3}$,
and the quantum fluctuations are neglected. Then, the effective 
potential of quark matter in $\beta$-equilibrium (with massless electrons) 
takes the form \cite{HS}:
\begin{eqnarray}
\Omega &=& \Omega_{0}
-\frac{1}{12\pi^2}\left(\mu_{e}^{4}+2\pi^{2}T^{2}\mu_{e}^{2}
+\frac{7\pi^{4}}{15} T^{4} \right) 
+\frac{\left(m-m^{(0)}\right)^2}{4G_S} \nonumber\\
&+&\frac{\Delta^2}{4G_D}
-\sum_{I} \int\frac{d^3 p}{(2\pi)^3} \left[E_{I}
+2 T\ln\left(1+e^{-E_{I}/T}\right)\right],
\label{pot}
\end{eqnarray}
where $\Omega_{0}$ is a constant added to make the pressure of the 
vacuum vanishing (the bag constant can also be included on the right
hand side if necessary). Here, while including electrons, we neglected 
the contribution of neutrinos. This properly describes the situation 
inside compact stars after the deleptonization had occurred, i.e., 
after the temperature of matter dropped so low that the neutrino mean 
free path became larger than a typical stellar size. In protoneutron 
stars this happens in less than a minute after their birth.

The sum in the second line of Eq.~(\ref{pot}) runs over all (6 quark 
and 6 antiquark) quasiparticles. The dispersion relations and the 
degeneracy factors of the quasiparticles read
\begin{eqnarray}
E_{ub}^{\pm} &=& E(p) \pm \mu_{ub} , \hspace{26.6mm} [\times 1]
\label{disp-ub} \\
E_{db}^{\pm} &=& E(p) \pm \mu_{db} , \hspace{26.8mm} [\times 1]
\label{disp-db}\\
E_{\Delta^{\pm}}^{\pm} &=& E_{\Delta}^{\pm}(p) \pm \delta \mu .
\hspace{26.5mm} [\times 2]
\label{2-degenerate}
\end{eqnarray}
These can be obtained in a rather straightforward way from the 
quark propagator. Here, the following shorthand notation was 
introduced:
\begin{eqnarray}
E(p) &\equiv& \sqrt{{\bf p}^2+m^2}, \\
E_{\Delta}^{\pm}(p) &\equiv&
\sqrt{[E(p) \pm \bar{\mu}]^2 +\Delta^2},\\
\bar{\mu} &\equiv&
\frac{\mu_{ur} +\mu_{dg}}{2}
=\frac{\mu_{ug}+\mu_{dr}}{2}
=\mu-\frac{\mu_{e}}{6}+\frac{\mu_{8}}{3}, \label{mu-bar}\\
\delta\mu &\equiv&
 \frac{\mu_{dg}-\mu_{ur}}{2}
=\frac{\mu_{dr}-\mu_{ug}}{2}
=\frac{\mu_{e}}{2}. \label{delta-mu}
\end{eqnarray}
The thermodynamic potential that determines the pressure of quark 
matter, $\Omega_{\rm phys} =-P$, is obtained from the effective 
potential $\Omega$ in Eq.~(\ref{pot}) after substituting the values 
of $\mu_{8}$, $\mu_{e}$, $m$ and $\Delta$ which solve the color and 
electric charge neutrality conditions, i.e.,
\begin{equation}
n_{8}\equiv
-\frac{\partial \Omega}{\partial \mu_{8}}=0, \quad \mbox{and} \quad
n_{Q}\equiv
-\frac{\partial \Omega}{\partial \mu_{e}}=0,
\label{Q=0}
\end{equation}
as well as the gap equations, i.e.,
\begin{equation}
\frac{\partial \Omega}{\partial m}=0, \quad \mbox{and} \quad
\frac{\partial \Omega}{\partial \Delta}=0.
\label{gap-eqs}
\end{equation}
Only a solution that satisfies Eqs.~(\ref{Q=0}) and (\ref{gap-eqs})
can correspond to a neutral ground state of quark matter. If there
exist several such solutions, it is the solution which gives the 
highest value of the pressure that gives the ground state.

\subsubsection{Three regimes in neutral matter}

In studying neutral two-flavor quark matter, it was found that 
there exist three qualitatively different dynamical regimes which
differ by the strength of diquark coupling, as well as by the 
properties of the ground state \cite{SH,HS}. 

The first regime corresponds to weak diquark coupling. In this case, 
a weak Cooper pairing is completely suppressed by the mismatch 
between the Fermi momenta of the up and down quarks. In other words,
the appearance of the 2SC phase is in conflict with the constraint 
of charge neutrality. The ground state of neutral matter 
corresponds to the normal phase in this regime. One should
note, though, that a much weaker spin-1 pairing between quarks of 
same flavor is not forbidden in such neutral matter. In fact, if the
temperature is sufficiently low, two independent spin-1 condensates, 
one made of up quarks and the other made of down quarks, are inevitable.

The other limiting case is the strongly coupled regime. It is clear
that, if the value of the diquark coupling is sufficiently large, the 
color condensation could be made as strong as needed to overcome a 
finite mismatch between the Fermi surfaces of pairing quarks. In 
this argument, it is taken into account that the charge neutrality 
constraint is only slightly affected by the value of the diquark 
coupling. This is the case in all models that have been studied so 
far. In this regime, the ground state is in the 2SC phase, and the 
neutrality has little effect. Here, it might be appropriate to 
mention that the unpaired blue up and blue down quarks of the 
2SC phase cannot produce any additional spin-1 condensate. The 
main reason is that their chemical potentials are badly mismatched.

\begin{figure}[ht]
\begin{minipage}[t]{0.48\textwidth}
\includegraphics[width=0.99\textwidth]{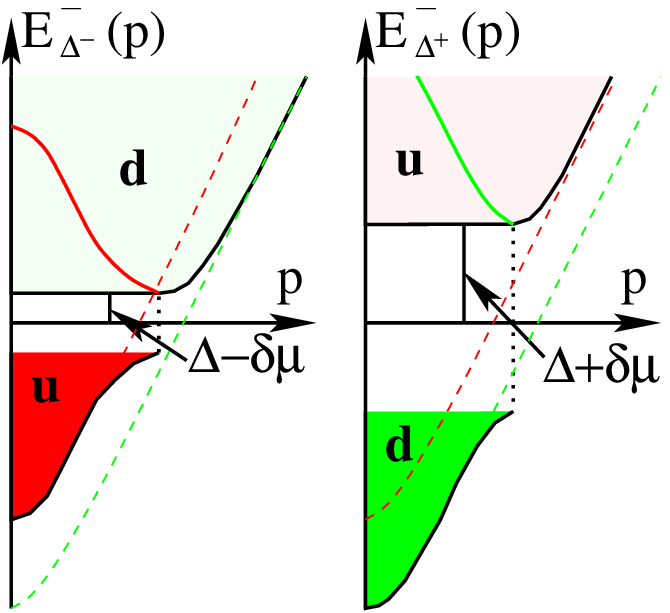}
\end{minipage}
\hspace{0.02\textwidth}
\begin{minipage}[t]{0.48\textwidth}
\includegraphics[width=0.99\textwidth]{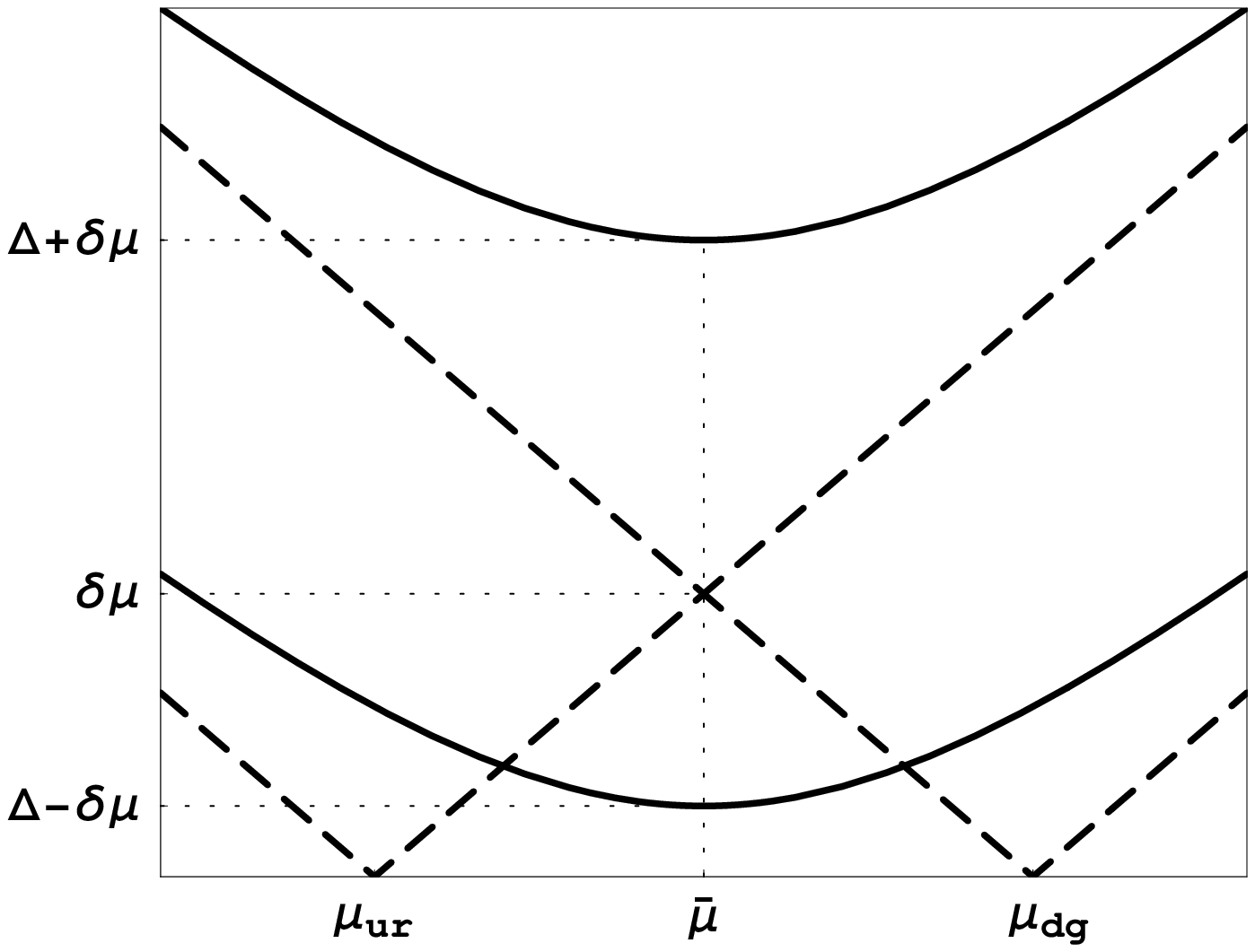}
\end{minipage}
\caption{The dispersion relations of quark quasiparticles in the 2SC 
phase (solid lines) and the normal quark matter phase (dashed lines). 
The low-energy part of the dispersion relations is shown in the 
right panel.}
\label{2sc-quasi}
\end{figure}

In Fig.~\ref{2sc-quasi} the dispersion relations of quasiparticles 
in the 2SC phase (i.e., in the strongly coupled regime) and in the 
normal quark matter phase (i.e., in the weakly coupled regime) are 
shown by the solid and dashed lines, respectively. The dispersion 
relations of unpaired blue quasiparticles are not shown. It should 
be clear, however, that they are similar to those in the normal 
phase, albeit slightly shifted in the horizontal direction. 

We see from Fig.~\ref{2sc-quasi} that there are two types of gapped 
quasiparticles in the spectrum of the 2SC phase. In fact, each of
them is a doublet with respect to the unbroken SU(2)$_c$ gauge group. 
One of the doublets has the gap $\Delta+\delta\mu$, and the other 
has the gap $\Delta-\delta\mu$, where $\delta\mu\equiv \mu_e/2$ and
$\Delta$ are the solutions of Eqs.~(\ref{Q=0}) and (\ref{gap-eqs})
that describe the ground state. When the mismatch parameter $\delta\mu$ 
is vanishing, two types of quasiparticles become degenerate. This
corresponds to the ideal 2SC phase, considered in the first lecture.

The actual values of the coupling constants, which define the 
strongly and weakly coupled dynamical regimes, can differ very 
much from one effective model of QCD to another. Thus, it is 
convenient to use the value of the following dimensionless ratio 
$\Delta/\delta\mu$ instead of the value of the diquark coupling
constant in a quantitative definition of such regimes. In this 
ratio, only the value of the gap $\Delta$ is strongly sensitive 
to the coupling constant. 

From our classification above, it should be clear that $\Delta/
\delta\mu=0$ corresponds to the weakly coupled regime. This 
should not be confused with the $G_D=0$ case. In fact, in any 
model, there should exist a critical value $G_D^{(1)}$ such that 
$\Delta/\delta\mu=0$ when $G_D<G_D^{(1)}$. As we shall see below,
the strongly coupled regime is given by $\Delta/\delta\mu>1$. 
This makes sense after noting that the values of the two gaps in 
the 2SC phase, $\Delta+\delta\mu$ and $\Delta-\delta\mu$, are 
positive when $\Delta/\delta\mu>1$ (see Fig.~\ref{2sc-quasi}). 

It turns out, that there also exists an intermediate regime, in which
the diquark coupling is neither too weak nor too strong. In this third
regime, the ground state is given by the so-called gapless 2SC phase 
\cite{SH,HS}. This phase corresponds to the ratio $\Delta/\delta\mu$ 
in the range from $0$ to $1$ (i.e., $0<\Delta/\delta\mu<1$). This regime 
will be discussed in more detail in the next subsection.

 \subsubsection{Gapless 2SC phase}

In the case of the NJL model used in Refs.~\cite{SH,HS}, the intermediate 
dynamical regime with $0<\Delta/\delta\mu<1$ is realized when the strength 
of the diquark coupling is given by $0.7\lesssim \eta \lesssim 0.8$
where $\eta\equiv G_D/G_S$. To be specific, below we use $\eta=0.75$.

Let us start from a brief discussion of the neutrality conditions in 
Eq.~(\ref{Q=0}). At zero temperature, they are satisfied approximately
when $\mu_8\approx 0$ and
\begin{eqnarray}
\mu_e & \approx & \frac{3}{5} \mu, \quad \mbox{for} \quad
\Delta > \delta\mu,
\label{Q=0_1}\\
\Delta & \approx & \sqrt{\frac{\mu_e^2}{4}-
\left(\frac{(\mu-2\mu_e/3)^3-\mu_e^3}
{6(\mu-\mu_e/6)^2}\right)^2} ,
\quad \mbox{for} \quad \Delta \leq \delta\mu.
\label{Q=0_2}
\end{eqnarray}
It should be noted that the solutions in the two regions of parameters,
$\Delta> \delta\mu$ and $\Delta \leq \delta\mu$, are very different.
One could check that this reflects a qualitatively different nature
of the quasiparticle spectra in the two regions. While all the modes
described by the dispersion relations in Eq.~(\ref{2-degenerate})
are gapped when $\Delta > \delta\mu$, there appear a gapless doublet 
mode when $\Delta \leq \delta\mu$.

\begin{figure}[ht]
\centerline{\includegraphics[width=8cm]{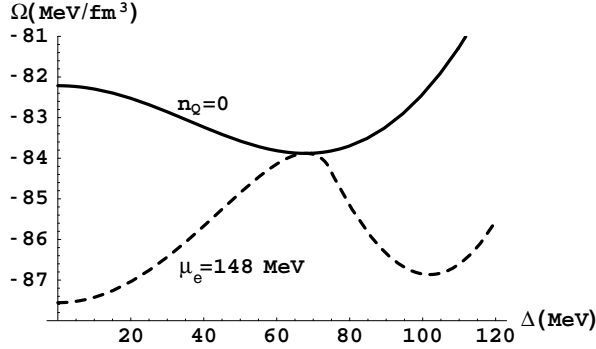}}
\caption{The effective potential as a function of the diquark gap
$\Delta$ calculated at a fixed value of the electron chemical
potential $\mu_e\approx 148$ MeV (dashed line), and the effective
potential defined along the neutrality line (solid line). The results
are plotted for $\mu=400$ MeV and $G_{D}=\eta G_{S}$ with $\eta=0.75$.}
\label{V2D}
\end{figure}

After imposing neutrality, the chemical potentials $\mu_e$ and $\mu_8$ 
are not free parameters, but (implicit) functions of $\Delta$. By making 
use of these functions $\mu_e=\mu_e(\Delta)$ and $\mu_8=\mu_8(\Delta)$
in Eq.~(\ref{pot}), one can define the effective potential for
neutral matter as a function of the gap parameter $\Delta$. This 
potential is shown graphically in Fig.~\ref{V2D} (solid line).

To emphasize the role of the neutrality condition, in
Fig.~\ref{V2D}, the result for the effective potential of non-neutral
quark matter at a fixed value of the electron chemical potential
$\mu_e\approx 148$ MeV is also shown (dashed line).
The chemical potential $\mu_e$ is chosen so that the electric charge
density vanishes when the value of the gap corresponds
to the maximum of the potential. This effective potential (labelled
``$\mu_e=148$ MeV'') describes negatively charged matter to
the left from the maximum, and positively charged matter to the
right from the maximum. On both sides, it is unphysical because
of a large (infinite in an infinite volume) contribution of the
Coulomb energy, see Eq.~(\ref{E_Coulomb}), that was neglected in 
Eq.~(\ref{pot}). It is fair to note that the charged minima of the 
dashed line may become physically important if mixed phases are 
allowed \cite{RR,SHH,mixed}.

\begin{figure}[ht]
\begin{minipage}[t]{0.48\textwidth}
\includegraphics[width=0.99\textwidth]{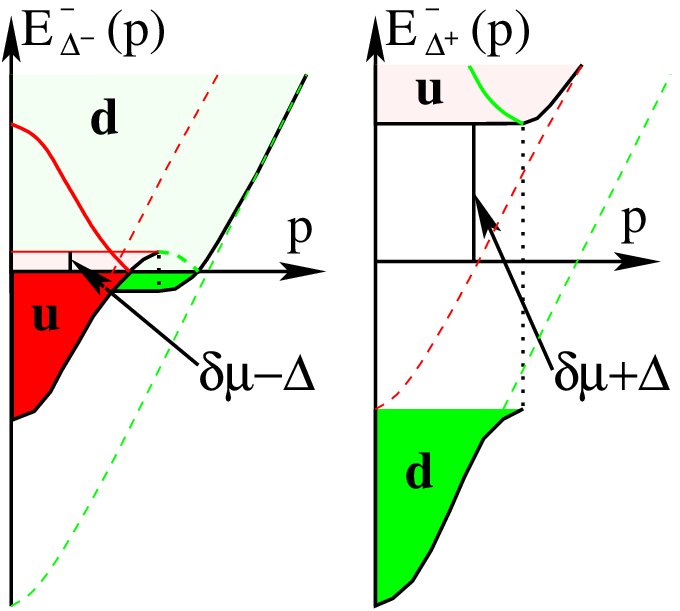}
\end{minipage}
\hspace{0.02\textwidth}
\begin{minipage}[t]{0.48\textwidth}
\includegraphics[width=0.99\textwidth]{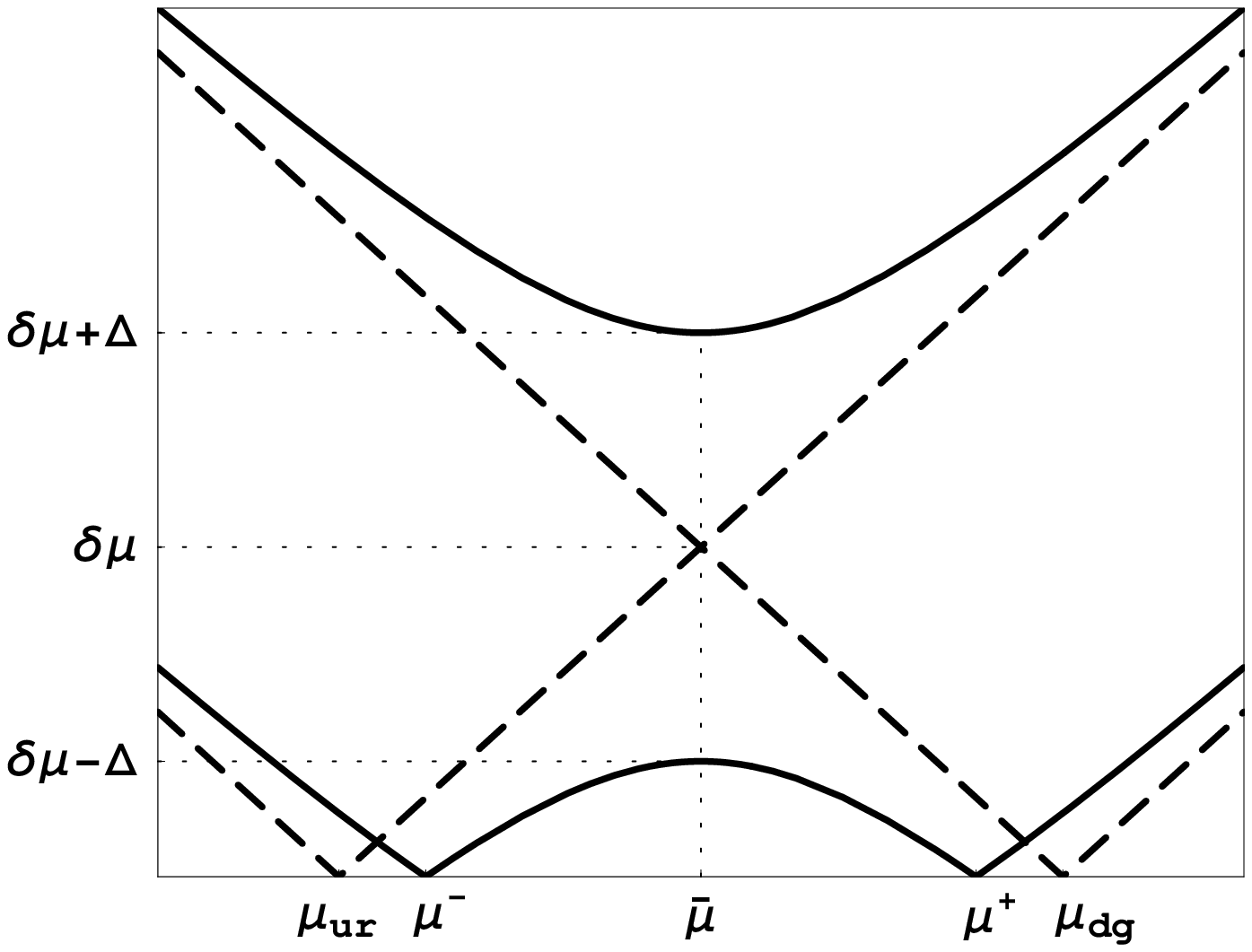}
\end{minipage}
\caption{The dispersion relations of quark quasiparticles in the g2SC 
phase (solid lines) and the normal quark matter phase (dashed lines). 
The low-energy part of the dispersion relations is shown in the 
right panel.}
\label{disp-rel}
\end{figure}

From the location of the minimum of the effective potential for
neutral quark matter (solid line in Fig.~\ref{V2D}), one determines
the value of the gap parameter in the ground state, $\Delta\approx
68$ MeV. It appears that $\Delta<\delta\mu\equiv \mu_e/2\approx 
74\,\mbox{MeV}$ in such a ground state, i.e., it corresponds to 
the intermediate regime in our classification. The low energy 
quasiparticle spectrum in this phase contains additional (as 
compared to the 2SC phase) gapless modes \cite{SH,HS}. This is
also seen clearly from the analytical expressions in 
Eqs.~(\ref{disp-ub})--(\ref{2-degenerate}). Graphically, the 
dispersion relations are shown in Fig.~\ref{disp-rel}. These
should be compared with the corresponding quasiparticle dispersion 
relations in the 2SC phase in Fig.~\ref{2sc-quasi}.

The most remarkable property of the quasiparticle spectrum in the
g2SC phase is that the low energy excitations ($E\ll \delta\mu-\Delta$)
are very similar to those in the normal phase. The only difference is
that the values of the chemical potentials of the free up and down
quarks are replaced by $\mu^{\pm}\equiv \bar{\mu}\pm
\sqrt{(\delta\mu)^2-\Delta^2}$. This simple observation may suggest
that the low energy (large distance scale) properties of the g2SC
phase are similar to those in the normal phase \cite{HS}.
For example, the Debye screening mass of the gluons of the SU$(2)_c$
subgroup should be nonzero. The latter, in fact, should be proportional
to the density of the gapless modes \cite{pi},
\begin{equation}
m_{1,D}^2\simeq \frac{2\alpha_s}{\pi}\left(
 \frac{(\mu^{-})^2\delta\mu}{\sqrt{(\delta\mu)^2-\Delta^2}}
+\frac{(\mu^{+})^2\delta\mu}{\sqrt{(\delta\mu)^2-\Delta^2}}
\right)\theta(\delta\mu-\Delta).
\label{m_D_1}
\end{equation}
Note that the corresponding value for the Debye screening mass in the 
ordinary 2SC phase is vanishing \cite{R-meissner}. The overall 
coefficient in Eq.~(\ref{m_D_1}) can be fixed by matching the value of 
the Debye screening mass with the known result in the normal phase. 
The Meissner screening mass of the gluons of the unbroken SU$(2)_c$ 
subgroup is vanishing, $m_{1,M}^2=0$, as it should be \cite{pi}.

The magnetic properties of the g2SC phase are nothing like those 
in the normal phase. The values of the Meissner screening masses 
of five gluons are {\em imaginary}, indicating a chromomagnetic 
instability in dense two-flavor quark matter \cite{pi}. This 
instability may lead to a gluon condensation, and possibly to
breaking the rotational or the translational symmetry in the 
true ground state of quark matter. This could be a completely new 
state, or this could be one of the states proposed earlier. For 
example, it may be a state with deformed quark Fermi surfaces 
\cite{MutherSed}, or it may turn 
into the Larkin-Ovchinnikov-Fulde-Ferrell state \cite{LOFF}, which 
was discussed in Ref.~\cite{loff-cs} in the context of quark matter. 
To make a certain statement, further studies in this direction are
needed.

\subsubsection{Finite temperature properties}

\begin{figure}[ht]
\begin{minipage}[t]{0.48\textwidth}
\includegraphics[width=0.99\textwidth]{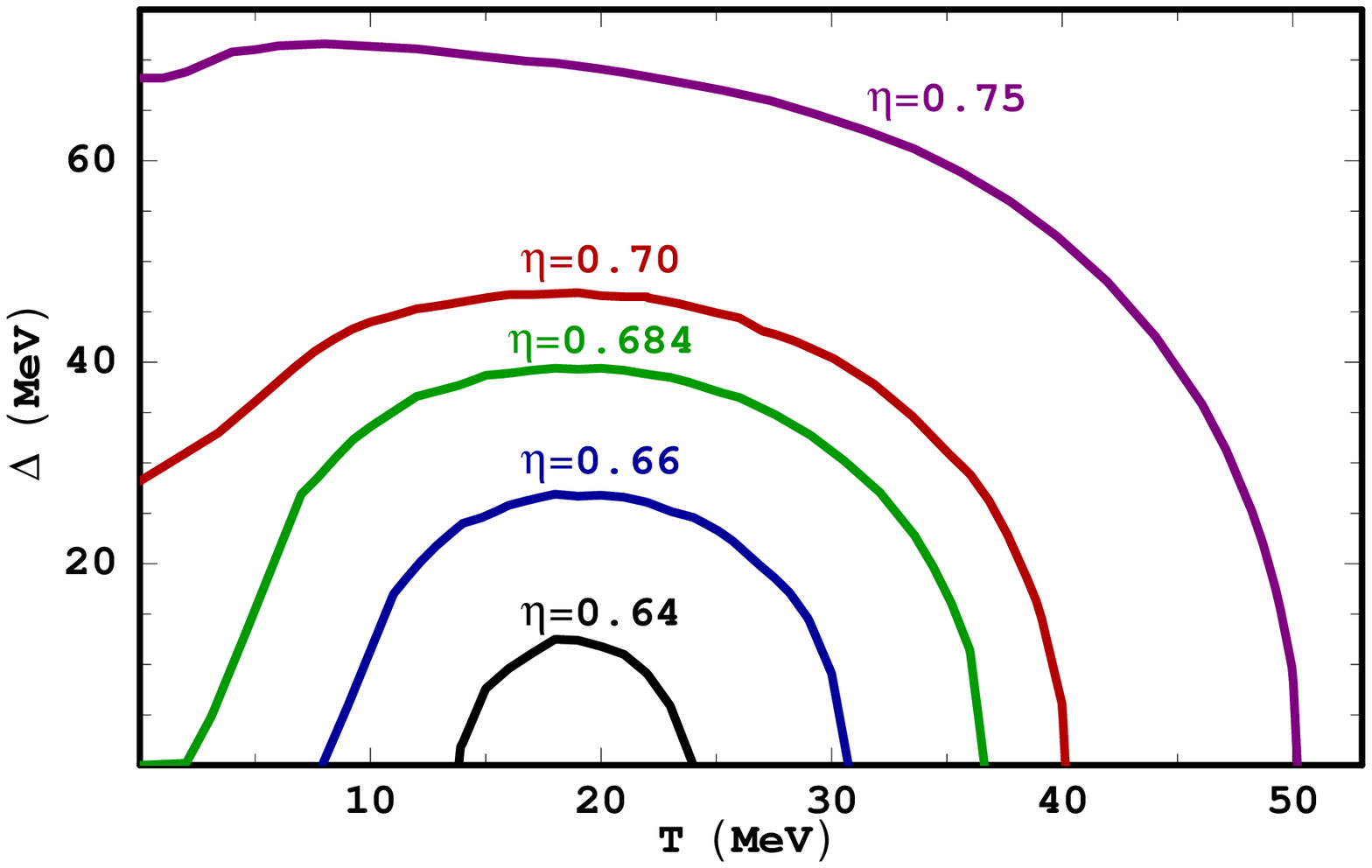}
\end{minipage}
\hspace{0.02\textwidth}
\begin{minipage}[t]{0.48\textwidth}
\includegraphics[width=0.99\textwidth]{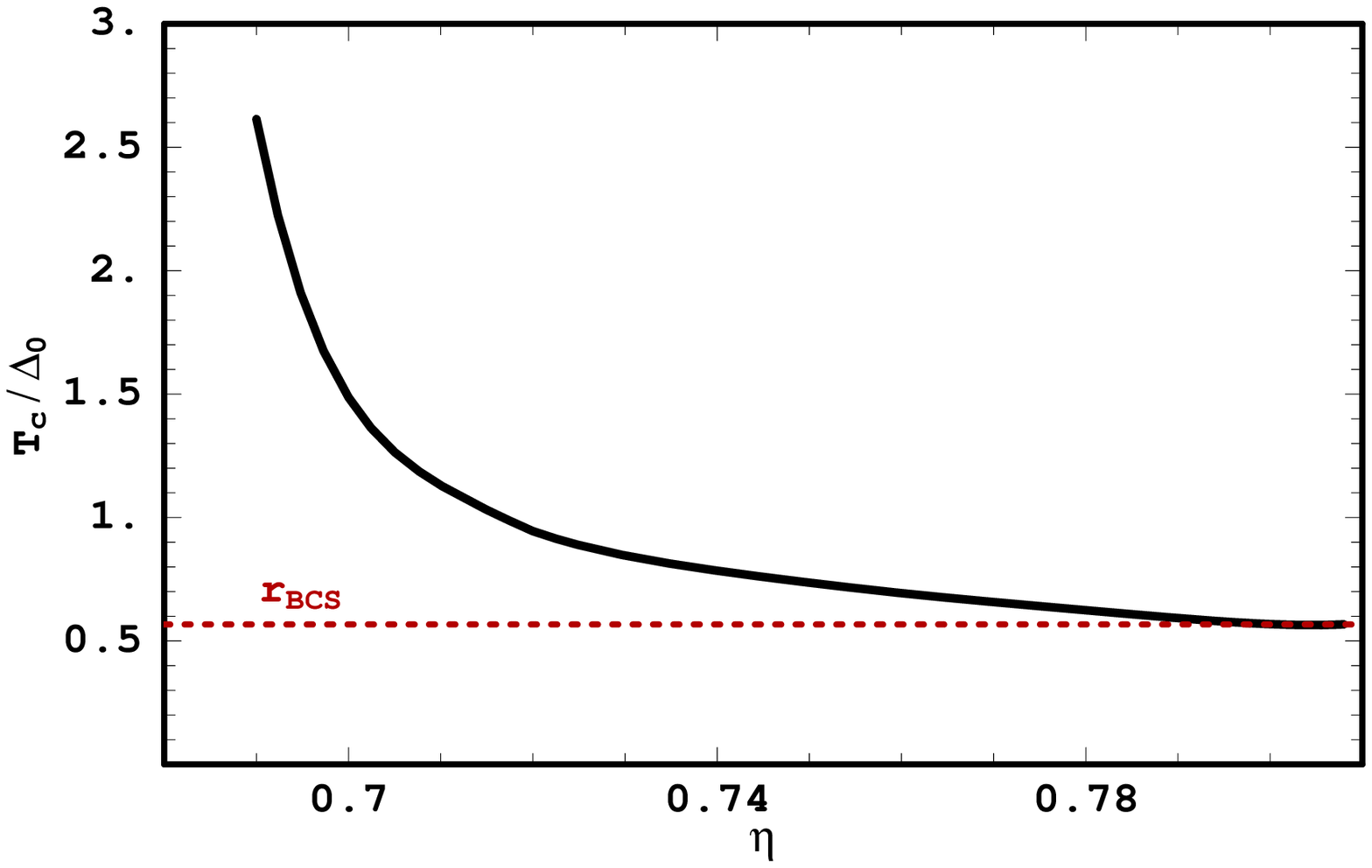}
\end{minipage}
\caption{The temperature dependence of the diquark gap in neutral
quark matter calculated for several values of the diquark coupling
strength $\eta=G_D/G_S$ (left panel), and the ratio of the critical 
temperature to the zero temperature gap in neutral quark matter as 
a function of the coupling strength $\eta$ (right panel).}
\label{gap-eta}
\end{figure}

The finite temperature properties of neutral quark matter in the
intermediate coupling regime are also very unusual \cite{HS,LZ}. 
The most striking are the results for the temperature dependence
of the gap parameter for various values of the diquark coupling
strengths. This is demonstrated by the plot in Fig.~\ref{gap-eta}.
As one can see, the results for not very strong coupling are very 
unusual. The gap function may take a finite value at nonzero 
temperatures even though
it is exactly zero at zero temperature. This possibility comes
about only because of the neutrality condition which introduces a 
mismatch between the Fermi momenta of pairing quarks. The thermal 
excitations smear the Fermi surfaces of the up and down quarks, and 
this opens the possibility of Cooper pairing which is forbidden 
at zero temperature. With increasing the temperature, the thermal 
fluctuations will eventually become too strong, and the pairing will 
be destroyed.

By looking at the results in the left panel of Fig.~\ref{gap-eta}, 
one should immediately realize that the value of the ratio of the 
critical temperature $T_c$ to the value of the gap at zero temperature 
$\Delta_0$ is not a universal number. In contrast to the 
Bardeen-Copper-Schrieffer theory, the ratio $T_c/\Delta_0$ 
in the neutral phase depends on the diquark coupling
constant. This is shown explicitly in the right panel in 
Fig.~\ref{gap-eta}. As we see, the ratio $T_c/\Delta_0$ can be 
arbitrarily large, and even remain strictly infinite for a 
range of coupling strengths.

\subsection{Gapless color-flavor locked phase}

While considering a realistic model of strange quark matter, one
should take into account that the strange quark mass is much larger 
than the masses of the light up and down quarks. Most likely, the 
actual value of strange quark mass in dense medium should be in 
the range between about $100\,\mbox{MeV}$ and $500\,\mbox{MeV}$. 
This is not negligible in comparison with a typical value of
the quark chemical potential $\mu\simeq 500\,\mbox{MeV}$ at 
densities existing in compact stars.

In application to the Cooper pairing dynamics responsible for color 
superconductivity, the main effect of a non-vanishing strange quark 
mass is a reduction of the strange quark Fermi momentum,
\begin{equation}
k_F^{(s)} = \sqrt{\mu^2-m_s^2} \simeq \mu-\frac{m_s^2}{2\mu}.
\label{k_F_shift}
\end{equation}
The magnitude of the reduction is approximately given by the value 
of $m_s^2/2\mu$. This quantity plays the role of a mismatch parameter
in three-flavor quark matter, which is similar to $\delta\mu\equiv 
\mu_e/2$ in two-flavor quark matter. For obvious reasons, this 
mismatch should interfere with Cooper pairing between strange 
and non-strange quarks. 

Because of a nonzero strange quark mass, the diquark condensate of 
the ideal CFL phase (\ref{LL-RR-cond}) may get distorted as follows
\cite{n_steiner,gCFL}:
\begin{equation}
\left\langle \psi_{L,i}^{a,\alpha} \epsilon_{\alpha\beta}
\psi_{L,j}^{b,\beta}\right\rangle
=-\left\langle \psi_{R,i}^{a,\dot\alpha} \epsilon_{\dot\alpha\dot\beta}
\psi_{R,j}^{b,\dot\beta}\right\rangle
\sim \Delta_1 \varepsilon_{ij1} \epsilon^{ab1}
  + \Delta_2 \varepsilon_{ij2} \epsilon^{ab2}
  + \Delta_3 \varepsilon_{ij3} \epsilon^{ab3} 
+\cdots,
\label{LL-RR-gCFL}
\end{equation}
where the ellipsis denote terms symmetric in color and flavor. 
As in the ideal CFL phase, they are small and not crucial for the
qualitative understanding of strange quark matter \cite{RSR}.

The three gap parameters $\Delta_1$, $\Delta_2$ and $\Delta_3$ in
Eq.~(\ref{LL-RR-gCFL}) describe down-strange, up-strange and up-down 
condensates of Cooper pairs, respectively. A nonzero value of 
$m_s^2/2\mu$ affects first and foremost the pairing between the 
strange and non-strange quarks, i.e., the gap parameters $\Delta_1$
and $\Delta_2$. Because of the color-flavor locking, preserved in 
the diquark condensate (\ref{LL-RR-gCFL}), this translates into 
a special role played by the blue quarks. 

By starting from the massless case and gradually increasing the value 
of strange quark mass, one finds that the CFL phase stays robust until 
a critical value of the control parameter $m_s^2/2\mu \simeq \Delta$ 
is reached \cite{gCFL}. Here, $\Delta$ is the gap parameter in the CFL
phase. (Strictly speaking, $\Delta\equiv \Delta_1=\Delta_2\approx\Delta_3$
in the CFL phase when $m_s\neq 0$.) Above the critical value, the 
charge neutrality cannot be accommodated in the CFL phase, and a new
gapless phase appears \cite{gCFL}.

A nice feature of the CFL phase is that it stays almost automatically 
electrically neutral \cite{enforce_n}. The reason is that Cooper 
pairing in the CFL phase helps to enforce the equal number densities of 
all three quark flavors, $n_u=n_d=n_s$. Since the sum of the charges 
of the up, down and strange quarks add up to zero, this insures that 
the electric charge density is vanishing, $n_Q=\frac{2}{3}n_u-\frac{1}{3}
n_d-\frac{1}{3}n_s=0$. This is exactly what happens in the CFL phase 
even at nonzero, but sub-critical values of the strange quark mass. 

It is the color rather than the electric charge neutrality that plays 
an important role in three-flavor quark matter at a nonzero value 
of $m_s^2/2\mu$. It has been mentioned above that, because of the
color-flavor locking, the blue quarks have a special status in 
the Cooper pairing dynamics. In order to avoid the violation of the 
color neutrality, then, a nonzero compensating value of the color 
chemical potential $\mu_8\simeq -m_s^2/2\mu$ is needed \cite{no2sc}. 
(Note that, in QCD with dynamical gluons, a nonzero value of the 
gluon condensate $\langle A_0^{8} \rangle$ will be generated instead 
\cite{DD}.) 

It was shown in Ref.~\cite{gCFL} that the CFL phase becomes gapless 
when $m_s^2/2\mu > \Delta$. In essence, the mechanism is the same as
in two-flavor quark matter, discussed earlier in this lecture. In 
order to get a slightly better understanding, it is instructive to 
consider how the strange quark mass interferes with the pairing 
between blue down quarks and green strange quarks 
\cite{gCFL}. After taking into account the shift of the strange 
quark Fermi momentum in Eq.~(\ref{k_F_shift}), the effective values 
of the chemical potentials of blue down and green strange quarks become
\begin{eqnarray}
\mu_{db} & \simeq & \mu - \frac{2}{3} \mu_8 
\simeq \mu + \frac{m_s^2}{3\mu},\\
\mu^{\rm eff}_{sg} & \simeq & \mu + \frac{1}{3} \mu_8 
- \frac{m_s^2}{2\mu}
\simeq \mu - \frac{2 m_s^2}{3\mu} ,
\end{eqnarray}
where we took into account that $\mu_8\simeq -m_s^2/2\mu$.
Up to an effective shift of the strange quark chemical potential, 
these expressions for the quark chemical potentials are similar 
to those in Eqs.~(\ref{mu-ur-ug})--(\ref{mu-db}) where $\mu_e=0$.
In analogy with the two-flavor case, see Eq.~(\ref{delta-mu}), 
we define
\begin{equation}
\delta\mu =\frac{\mu_{db}-\mu^{\rm eff}_{sg}}{2} = \frac{m_s^2}{2\mu},
\end{equation}
as a formal value of the mismatch parameter. As soon as $\delta\mu
>\Delta$, one should get the gapless modes in the quasiparticle
spectrum. Their dispersion relations are given by the same expression
(\ref{2-degenerate}), except that $\bar\mu \to 
(\mu_{db}+\mu^{\rm eff}_{sg})/2$.

\begin{figure}[ht]
  \includegraphics[width=0.9\textwidth]{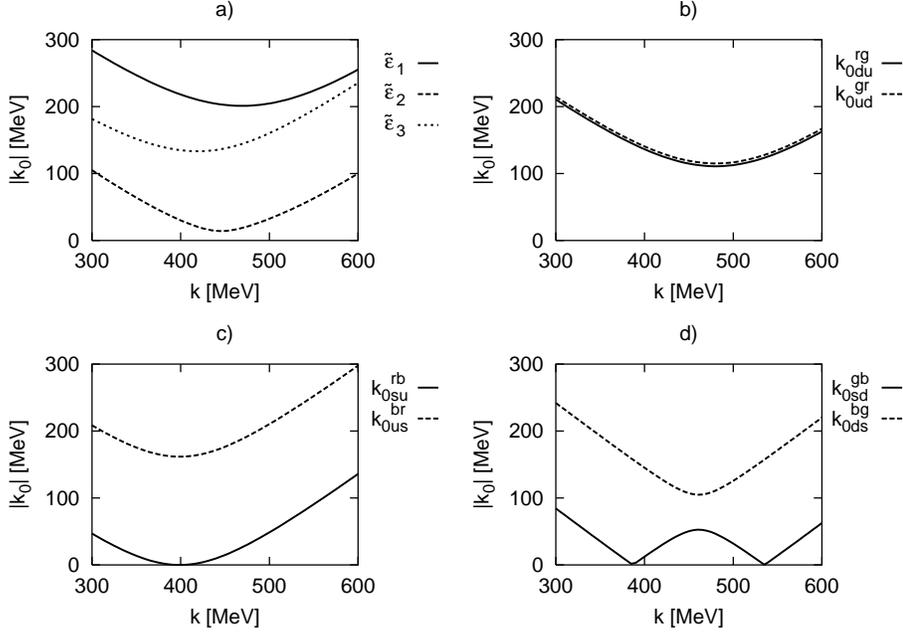}
  \caption{The quasiparticle dispersion relations for electric 
  and color neutral color superconducting quark matter at $T=0$, 
  $\mu=500\,\mbox{MeV}$, and $m_s=400\,\mbox{MeV}$, see Ref.~\cite{RSR}.}
  \label{disp-numer}
\end{figure}

A typical result for the low-energy dispersion relations of all nine 
quasiparticle are shown in Fig.~\ref{disp-numer} which is taken 
from Ref.~\cite{RSR}. It should be noted that the two dispersion 
relations in panel (d) are qualitatively the same as those in 
Fig.~\ref{disp-rel} in the gapless 2SC case.

Physical properties of the gCFL phase are very different from 
those of the CFL phase. The presence of gapless quasiparticle modes
with a density of states proportional to $\mu^2$ has a large effect 
on the thermodynamics as well as on the transport. In contrast to 
the CFL phase which is an insulator, the gCFL phase is a metal 
with a nonzero number density of electrons. Also, the neutrino 
emissivity rate from the gCFL phase should be rather high.
It is dominated by the $\beta$-processes: 
$\tilde{d}_b \to \tilde{u}_b + e^{-} + \bar{\nu}_{e}$ and 
$\tilde{u}_b + e^{-} \to \tilde{d}_b+\nu_{e}$, where $\tilde{u}_b$ 
and $\tilde{d}_b$ denote the gapless quasiparticles whose dispersion
relations are represented by the solid lines in panels (c) and (d) in 
Fig.~\ref{disp-numer}. Relatively large contributions to the neutrino 
emissivity may also come from the processes with additional gluons: 
$\tilde{s}_{r,g} \to \tilde{u}_b + e^{-} + \bar{\nu}_{e}+g_{4,5,6,7}$ and
$\tilde{u}_b + e^{-} \to \tilde{s}_{r,g} + \nu_{e} + g_{4,5,6,7}$, where 
$g_{4,5,6,7}$ denotes one of the gluons described by the gauge fields 
$A^{4}_{\mu}$, $A^{5}_{\mu}$, $A^{6}_{\mu}$ or $A^{7}_{\mu}$.

\subsection{Mixed phases with color superconducting components}

In addition to homogeneous (one-component) phases, one could also 
study various mixed phases of dense quark matter. The neutrality 
in such phases is imposed not locally, but {\it globally} (i.e., 
on average). For example, one can construct a phase which is made 
of alternating layers of different coexisting components. While 
each of the component may have a nonzero charge density, the 
average charge density of the whole construction is vanishing. 
The Coulomb energy in a mixed phase is nonzero, but it does not 
grow with the volume of the system. There 
is also a contribution to the total energy that comes from surfaces 
separating components of the mixed phase \cite{RR,SHH,mixed}. If 
the corresponding surface tension is sufficiently large, the mixed 
phase is disfavored energetically, and therefore it cannot appear. 
If, on the other hand, the surface tension is small, the mixed phase 
can be the ground state of matter, replacing a homogeneous gapless 
phase. 

 \subsubsection{Gibbs construction}

\begin{figure}[ht]
\includegraphics[width=0.6\textwidth]{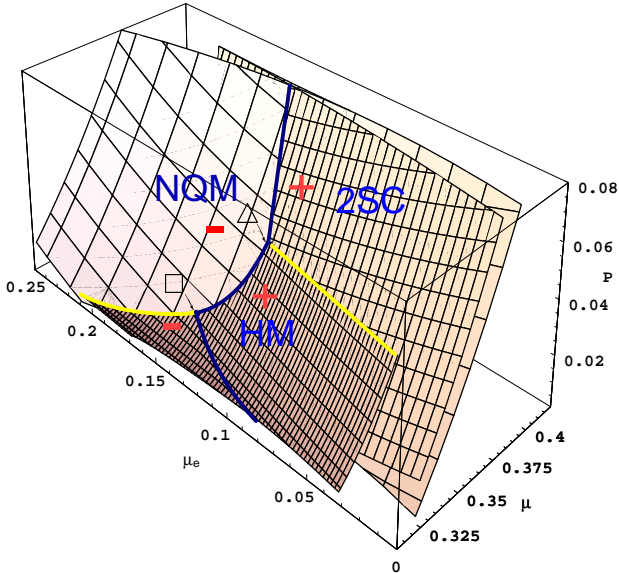}
\caption{\label{fig-front-quark}
Pressure as a function of $\mu\equiv\mu_B/3$ and $\mu_e$ for the
hadronic phase (at the bottom), for the 2SC phase (on the right in 
front) and the normal quark phase (on the left). The dark thick line 
follows the neutrality line in the hadronic phase, and in two mixed 
phases: (i) the mixed phase of hadronic and normal quark matter; and 
(ii) the mixed phase of normal and color superconducting quark matter.}
\end{figure}

Here is a brief introduction to the general method of constructing 
mixed phases by imposing the Gibbs conditions of equilibrium 
\cite{glen92,Weber}. From the physical point of view, the Gibbs
conditions enforce the mechanical as well as chemical equilibrium 
between different components in a mixed phase. This is achieved by 
requiring that the pressure of different components inside the mixed 
phase are equal, and that the chemical potentials ($\mu$ and $\mu_{e}$) 
are the same across the whole mixed phase. For example, in application 
to the mixed phase made of the normal and the 2SC quark components 
\cite{SHH}, these conditions read
\begin{eqnarray}
P^{(NQ)}(\mu,\mu_{e}) &=& P^{(2SC)}(\mu,\mu_{e}),
\label{P=P}\\
\mu &=& \mu^{(NQ)}=\mu^{(2SC)},
\label{mu=mu}\\
\mu_{e} &=& \mu^{(NQ)}_{e}=\mu^{(2SC)}_{e}.
\label{mue=mue}
\end{eqnarray}
It is easy to visualize these conditions by plotting the pressure 
surfaces as functions of chemical potentials ($\mu$ and $\mu_{e}$) 
for the components of the mixed phase. Graphically, this is shown 
in Fig. \ref{fig-front-quark}. The Gibbs conditions are automatically 
satisfied along the intersection line of the pressure surfaces (dark 
solid line in Fig.~\ref{fig-front-quark}).

Different components of the mixed phase occupy different volume 
fractions in space. To describe this quantitatively, it is convenient 
to introduce the volume fraction of each component. As an example, 
let us consider the quark mixed phase 
made of the normal and the 2SC component \cite{SHH}. We denote the 
volume fraction of the normal phase by $\chi^{NQ}_{2SC}\equiv V_{NQ}/V$
(in general, the notation $\chi^{A}_{B}$ means ``the volume fraction of 
phase A in a mixture with phase B"). Then, the volume fraction of the 
2SC phase is given by $\chi^{2SC}_{NQ}=(1-\chi^{NQ}_{2SC})$. From the 
definition, it is clear that $0\leq \chi^{A}_{B} \leq 1$.

The average electric charge density of the mixed phase is determined by
the charge densities of its components taken in the proportion of the
corresponding volume fractions. Thus,
\begin{equation}
n^{(MP)}_{e}(\mu,\mu_e) = \chi^{NQ}_{2SC} n^{(NQ)}_{e}(\mu,\mu_e)
+(1-\chi^{NQ}_{2SC}) n^{(2SC)}_{e}(\mu,\mu_e).
\end{equation}
If the charge densities of the two components have opposite signs, one
can impose the global charge neutrality condition, $n^{(MP)}_{e}=0$.
Otherwise, a neutral mixed phase could not exist. In the case of
quark matter, the charge density of the normal quark phase is negative,
while the charge density of the 2SC phase is positive along the line of
the Gibbs construction (dark solid line in Fig. \ref{fig-front-quark}).
Therefore, a neutral mixed phase exists. The volume fractions of its
components are
\begin{eqnarray}
\chi^{NQ}_{2SC} &=& \frac{n^{(2SC)}_{e}}{n^{(2SC)}_{e}-n^{(NQ)}_{e}}, \\
\chi^{2SC}_{NQ} &\equiv& 1-\chi^{NQ}_{2SC}=
\frac{n^{(NQ)}_{e}}{n^{(NQ)}_{e}-n^{(2SC)}_{e}}.
\end{eqnarray}

After the volume fractions have been determined from the condition of the
global charge neutrality, the energy density of the corresponding mixed 
phase can also be calculated,
\begin{equation}
\varepsilon^{(MP)}(\mu,\mu_e) = \chi^{NQ}_{2SC} \varepsilon^{(NQ)}(\mu,\mu_e)
+(1-\chi^{NQ}_{2SC}) \varepsilon^{(2SC)}(\mu,\mu_e).
\end{equation}
This is essentially all what one needs in order to construct the equation 
of state of the mixed phase.

 \subsubsection{Surface tension and Coulomb forces}

In the above construction, the effects of the Coulomb forces and the
surface tension between different components of the mixed phase were 
neglected. 
In reality, these might be important. In particular, the balance
between the Coulomb forces and the surface tension determines the
geometries of different components inside the mixed phase.

In the mixed phase made of the normal and 2SC components, for example,
nearly equal volume fractions of the two quark phases are likely to 
form alternating layers (slabs) of matter. The energy cost per unit 
volume to produce such layers scales as $\sigma^{2/3}(n_{e}^{(2SC)}
-n_{e}^{(NQ)})^{2/3}$ where $\sigma$ is the surface tension 
\cite{geometry}. Therefore, the quark mixed phase is a favorable 
phase of matter only if the surface tension is not too large. The 
estimates of Ref.~\cite{RR} show that the value of $\sigma$ at the
boundary between the normal and the 2SC phases is only about 
$5$ to $10\,\mbox{MeV/fm}^{2}$. The maximum value of $\sigma$
allowed in the mixed phase, on the other hand, is of order 
$10$ to $15\,\mbox{MeV/fm}^{2}$ 
when the value of the quark chemical potential is in the range between 
$400$ and $500\,\mbox{MeV}$. Therefore, one should conclude that the 
mixed phase is more favorable than the gapless 2SC phase. 

Here, it is fair to mention that the conclusion of Ref.~\cite{RR} 
may not be final yet. The reason is that the thickness of the layers, 
which scales as $\sigma^{1/3} (n_{e}^{(2SC)}-n_{e}^{(NQ)})^{-2/3}$ 
\cite{geometry}, is estimated to be of order $5$ to $10\,\mbox{fm}$ 
in the mixed phase \cite{RR,SHH}. This is comparable to the value of 
the Debye screening length in each of the two quark phases. Therefore, 
the effects of charge screening, which have been neglected so far and 
which are known to increase the surface energy \cite{screening}, may 
still change the conclusion. 

\subsection{Summary}

In recent years there was a lot of progress in understanding the 
mechanism of color superconductivity and in studying physical 
properties of various possible color superconducting phases of dense 
quark matter. At present, one can say that the QCD theory of strong 
interactions at asymptotic densities is most likely to have a color 
superconducting ground state. Unfortunately, however, one cannot 
say with certainty what is the critical value of the baryon density 
at which color superconductivity first appears. Because of this, one
does not know for sure if color superconducting phases could 
appear at highest densities existing in the Universe. 

In nature, very dense baryonic matter exists in compact stars.
Thus, it is possible that color superconductivity exists in the central 
regions of the heaviest compact stars. This conjecture has not yet 
been (dis-)proved, and it should come under scrutiny in the near 
future. In this connection, it might be appropriate to mention that 
there were recent observations on the cooling \cite{quark1} of one 
neutron star and the radius \cite{quark2} of another that led the 
authors to suggest the presence of quark matter. These suggestions 
have been disputed \cite{no-quark1,no-quark2}. However, the current 
situation, where there is no unambiguous evidence that deconfined 
quarks play a role in compact stars, may well change as further 
observations are made.

One of the main consequences of color superconductivity in dense 
matter is the appearance of an energy gap in the quasiparticle 
spectrum. In QCD, the typical value of the gap is $100\,\mbox{MeV}$. 
The presence of such a large gap is likely to affect many physical 
properties of matter, which in their turn should be reflected in the
observational properties of stars. It is of great interest, 
therefore, to perform a systematic study of the properties of dense 
baryonic matter under conditions typical for the interior of stars. 
The goal of such studies should be a clear theoretical prediction of 
the most favorable phases of matter inside stars. Also, one 
should work out the key physical properties of different phases 
that could have an unambiguous imprint on the observational data 
from stars. Such a theoretical basis would be crucial in an attempt
to discover new (e.g., quark) states of matter in nature.

The current understanding of color superconducting phases of QCD at 
large baryon chemical potential $\mu$ may also turn out to be useful 
for developing alternative methods to treat QCD at nonzero $\mu$ on the 
lattice \cite{HH,Hands}. As is known, there are not so many methods 
to study QCD from first principles. One of them is the
method of lattice calculations \cite{lattice}. As of now, the lattice 
calculations have produced many results for QCD at finite temperatures 
and vanishing baryon chemical potential $\mu=0$ \cite{lattice-res}. 
There are also some results at nonzero but small values of the baryon 
chemical potential \cite{lattice-mu}. The lattice calculations at large 
nonzero $\mu$ are problematic because of the famous sign problem.
At $\mu\neq 0$, 
the fermion determinant in the functional integral becomes complex,
and the standard numerical techniques, which rely on a probability 
interpretation of the integrand, cannot be easily applied. Thus, 
new approaches and new ideas are needed.

Building on the knowledge gained from lattice calculations at nonzero 
temperature as well as from the effective models of QCD at nonzero 
chemical potential, one would like to get eventually a complete QCD 
phase diagram in the plane of temperature and quark chemical potential.
The knowledge of such a diagram is not only of a theoretical importance.
Different regions of the phase diagram describe physical conditions 
that can be realized, for example, during the evolution of the early 
Universe or in the heavy ion collisions. There is also a region at 
nonzero baryon chemical potentials which is responsible for the 
physics of stars. In order to understand the corresponding physics, 
one first has to know the locations of the critical lines and 
the types of the phase transitions. Some basic features of the QCD 
diagram have already started to emerge \cite{RSR,FKR,Halasz,Rajpd}. 
It will probably take many years before the complete phase diagram 
is understood in detail.

\subsection*{Acknowledgements}
I would like to thank the organizers of the IARD 2004 conference and 
the organizers of the Helmholtz International Summer School and 
Workshop on {\sl Hot points in Astrophysics and Cosmology} for their
invitation to present these lectures. Also, I thank
Michael Buballa, Valery Gusynin, Paul Ellis, Matthias Hanauske, 
Deog-Ki Hong, Mei Huang, Volodya Miransky, Micaela Oertel, Dirk Rischke, 
Stefan R\"{u}ster, Gordon Semenoff, and Rohana Wijewardhana for very 
fruitful collaboration on various topics in color superconductivity. 
This work was supported by Gesellschaft f\"{u}r Schwerionenforschung 
(GSI) and by Bundesministerium f\"{u}r Bildung und Forschung (BMBF).

\end{document}